\documentclass[a4paper]{iopart}
\usepackage{iopams}
\usepackage{graphicx}
\usepackage[T1]{fontenc}
\usepackage{eurosym}
\usepackage{rotating}
\usepackage[dvips]{color}

\newcommand{\be}{\begin{equation}}
\newcommand{\ee}{\end{equation}}
\newcommand{\delCP}{\ensuremath{\delta_{\rm CP}}}
\newcommand{\nubarmu}{\ensuremath{\bar{\nu}_{\mu}}}
\newcommand{\stheta}{\sin^22\theta_{13}}
\newcommand{\thetaot}{\ensuremath{\theta_{13}}\,}
\newcommand{\nue}{\ensuremath{\nu_{e}}}
\newcommand{\nubare}{\ensuremath{\bar{\nu}_{e}}}
\newcommand{\He}{\ensuremath{^6{\mathrm{He}\,}}}
\newcommand{\Ne}{\ensuremath{^{18}{\mathrm{Ne}\,}}}
\newcommand{\numu}{\ensuremath{\nu_{\mu}}}
\newcommand{\anue}{\overline{{\mathrm\nu}}_{\mathrm e}}
\newcommand{\anumu}{\overline{{\mathrm\nu}}_{\mathrm \mu}}

\newcommand{\nunubar}[1]{\mbox{\raisebox{0ex}{$\stackrel{\scriptscriptstyle (-)}{\displaystyle \nu_#1}$}}}
\newcommand{\WC}{water Cherenkov}

\newcommand{\flux}{\mbox{$ \mathrm{cm}^{-2}~\mathrm{s}^{-1}$}}

\newcommand{\refTab}[1]{Tab.~\ref{#1}}
\newcommand{\refFig}[1]{Fig.~\ref{#1}}
\newcommand{\refSec}[1]{Sec.~\ref{#1}}

\begin{document}
\bibliographystyle{Campagne}

\title[Large underground, liquid based detectors for astro-particle physics in Europe]{Large underground, liquid based detectors for astro-particle physics in Europe: scientific case and prospects}
\author{
D~Autiero~$^1$,
J~\"Ayst\"o~$^2$,
A~Badertscher~$^3$,
L~Bezrukov~$^4$,
J~Bouchez~$^5$,
A~Bueno~$^6$,
J~Busto~$^7$,
J-E~Campagne~$^8$,
Ch~Cavata~$^9$,
L~Chaussard~$^1$,
A~de~Bellefon~$^{10}$,
Y~Déclais~$^1$, 
J~Dumarchez~$^{11}$,
J~Ebert~$^{12}$,
T~Enqvist~$^{13}$,
A~Ereditato~$^{14}$,
F~von~Feilitzsch~$^{15}$,
P~Fileviez~Perez~$^{16}$,
M~G\"oger-Neff~$^{17}$,
S~Gninenko~$^4$,
W~Gruber~$^3$,
C~Hagner~$^{12}$,
M~Hess~$^{14}$,
K~A~Hochmuth~$^{17}$,
J~Kisiel~$^{18}$,
L~Knecht~$^3$,	
I~Kreslo~$^{14}$,
V~A~Kudryavtsev~$^{19}$,
P~Kuusiniemi~$^{13}$,
T~Lachenmaier~$^{15}$,
M~Laffranchi~$^3$,
B~Lefievre~$^10$,
P~K~Lightfoot~$^{19}$,
M~Lindner~$^{20}$,
J~Maalampi~$^2$,
M~Maltoni~$^{21}$,
A~Marchionni~$^3$,
T~Marrodán~Undagoitia~$^{15}$,
J~Marteau~$^1$,
A~Meregaglia~$^3$,
M~Messina~$^{14}$,
M~Mezzetto~$^{22}$,
A~Mirizzi~$^{17,23}$,
L~Mosca~$^9$,
U~Moser~$^{14}$,
A~Müller~$^3$,
G~Natterer~$^3$,
L~Oberauer~$^{15}$,
P~Otiougova~$^3$,
T~Patzak~$^{10}$,
J~Peltoniemi~$^{13}$,
W~Potzel~$^{15}$,
C~Pistillo~$^{14}$,
G~G~Raffelt~$^{17}$,
E~Rondio~$^{24}$,
M~Roos~$^{25}$,
B~Rossi~$^{14}$,
A~Rubbia~$^3$,
N~Savvinov~$^{14}$,
T~Schwetz~$^{26}$,
J~Sobczyk~$^{27}$,
N~J~C~Spooner~$^{19}$,
D~Stefan~$^{28}$,
A~Tonazzo~$^{10}$,
W~Trzaska~$^2$,
J~Ulbricht~$^3$,
C~Volpe~$^{29}$,
J~Winter~$^{15}$,
M~Wurm~$^{15}$,
A~Zalewska~$^{28}$
and
R~Zimmermann~$^{12}$
}
\address{$^1$ IPNL, Université Claude Bernard Lyon 1, CNRS/IN2P3, 69622 Villeurbanne, France}
\address{$^2$ Department of Physics, University of Jyv\"askyl\"a, Finland}
\address{$^3$ Institut f\"{u}r Teilchenphysik,  ETHZ, Z\"{u}rich, Switzerland}
\address{$^4$ Institute for Nuclear Research, Russian Academy of Sciences, Moscow, Russia}
\address{$^5$ CEA - Saclay, Gif sur Yvette and APC Paris, France}
\address{$^6$ Dpto Fisica Teorica y del Cosmos \& C.A.F.P.E., Universidad de Granada, Spain}
\address{$^7$ Centre de Physique des Particules de Marseille (CPPM), IN2P3-CNRS et Université d'Aix-Marseille II, Marseille, France}
\address{$^8$ LAL, Université Paris-Sud, IN2P3/CNRS, Orsay, France}
\address{$^9$ CEA - Saclay, Gif sur Yvette, France}
\address{$^{10}$ Astroparticule et Cosmologie (APC), CNRS, Université Paris VII, CEA, Observatoire de Paris, Paris, France}
\address{$^{11}$ Laboratoire de Physique Nucléaire et des Hautes Energies (LPNHE), IN2P3-CNRS et Universités Paris VI et Paris VII, Paris, France}
\address{$^{12}$ Universität Hamburg, Institut für Experimentalphysik, Hamburg, Germany}
\address{$^{13}$ CUPP, University of Oulu, Finland}
\address{$^{14}$ Laboratorium f\"{u}r  Hochenergie Physik, Bern Universit\"at, Bern, Switzerland}
\address{$^{15}$ Technische Universit\"at M\"unchen, Physik-Department E15, Garching, Germany}
\address{$^{16}$ Centro de Fisica Teorica de Particulas, Instituto Superior Tecnico, Departamento de Fisica, Lisboa, Portugal}
\address{$^{17}$ Max-Planck-Institut f\"ur Physik (Werner-Heisenberg-Institut), M\"unchen, Germany}
\address{$^{18}$ Institute of Physics, University of Silesia, Katowice, Poland}
\address{$^{19}$ Department of Physics and Astronomy, University of Sheffield, Sheffield, United Kingdom}
\address{$^{20}$ Max-Planck-Institut fuer Kernphysik, Heidelberg, Germany}
\address{$^{21}$ Departamento de F\'{\i}sica Te\'orica \& Instituto de F\'{\i}sica
Te\'orica, Facultad de Ciencias C-XI, Universidad Aut\'onoma de Madrid, Cantoblanco, Madrid, Spain}
\address{$^{22}$ INFN Sezione di Padova, Padova, Italy}
\address{$^{23}$ INFN Sezione di Bari and Dipartimento di Fisica, Università di Bari, Bari, Italy}
\address{$^{24}$ A. Soltan Institute for Nuclear Studies, Warsaw, Poland}
\address{$^{25}$ Department of Physical Sciences, University of Helsinki, Finland}
\address{$^{26}$ CERN, Physics Department, Theory Division, Geneva, Switzerland}
\address{$^{27}$ Institute of Theoretical Physics, Wroclaw University, Wroclaw, Poland}
\address{$^{28}$ H. Niewodniczanski Institute of Nuclear Physics, Krakow, Poland}
\address{$^{29}$ Institut de Physique Nucleaire d'Orsay (IPNO), Groupe de Physique Theorique, Université de Paris-Sud XI, Orsay, France}
\ead{campagne@lal.in2p3.fr}


\begin{abstract}

This document reports on a series of experimental and theoretical studies conducted to 
assess the astro-particle physics potential of three future large-scale particle detectors 
proposed in Europe as next generation underground observatories. 
The proposed apparatus employ three different and, to some extent, complementary detection techniques: 
GLACIER (liquid Argon TPC), LENA (liquid scintillator) and MEMPHYS (\WC), based on the use of large mass of liquids
as active detection media.
The results of these studies are presented along with a critical discussion of the performance attainable by the three proposed 
approaches coupled to existing or planned underground laboratories, 
in relation to open and outstanding physics issues such as the search for matter instability, the detection
of astrophysical- and geo-neutrinos and to the possible use of these detectors in future high-intensity
neutrino beams.\\

\end{abstract}


\submitto{Journal of Cosmology and Astroparticle Physics}

\maketitle

\section{Physics motivation}
\label{sec:Phys-Intro}

Several outstanding physics goals could be achieved by the next generation of large underground observatories
in the domain of astro-particle and particle physics, neutrino astronomy and cosmology.
Proton decay \cite{Pati:1973rp}, in particular, is one of the most exciting prediction of Grand Unified Theories 
(for a review see \cite{Nath:2006ut}) aiming at the 
unification of fundamental forces in Nature. It remains today one of the most relevant open questions
of particle physics. Its discovery would certainly represent a fundamental milestone, contributing to clarifying our
understanding of the past and future evolution of the Universe.  

Several experiments have been built and conducted to search for proton decay but they only yielded lower limits to the proton lifetime. 
The window between the predicted proton lifetime (in the simplest models typically below $10^{37} $ years) and that excluded 
 by experiments \cite{Kobayashi:2005pe}
($O$($10^{33}$) years, depending on the channel) is within reach, 
and the demand to fill the gap grows with the progress in other domains of particle physics, astro-particle physics and cosmology. 
To some extent, also a negative result from next generation high-sensitivity experiments
would be relevant to rule-out some of the
theoretical models based on SU(5) and SO(10) gauge symmetry or to further constrain the range of allowed parameters.
Identifying unambiguously proton decay and measuring its lifetime would set a firm scale for any Unified Theory, narrowing
the phase space for possible models and their parameters. This will be a mandatory step to go forward 
beyond the Standard Model of elementary particles and interactions.

Another important physics subject is the physics of 
astrophysical
neutrinos, as those from supernovae, from the Sun and from the interaction of primary cosmic-rays with the Earth's atmosphere. Neutrinos are above all important messengers from stars. 
Neutrino astronomy has a glorious although recent history, from the detection of solar neutrinos 
 \cite{Davis:1968cp,Hirata:1989zj,Anselmann:1992um,Abdurashitov:1994bc,Smy:2002rz,Aharmim:2005gt,Altmann:2005ix} 
to the observation of neutrinos from supernova explosion, \cite{Hirata:1987hu,Bionta:1987qt,Alekseev:1988gp},
acknowledged by the Nobel Prizes awarded to M. Koshiba and R. Davis.
These observations have given valuable information for a better understanding of the functioning
of stars and of the properties of neutrinos. However, much more information could be obtained if the energy spectra of
stellar neutrinos were known with higher accuracy.
Specific neutrino observations could give detailed information on the conditions of the production zone, 
whether in the Sun or in a supernova. 
A supernova explosion in our galaxy would be extremely important as the evolution mechanism of the collapsed star 
is still a puzzle for astrophysics.
An even more fascinating challenge would be observing neutrinos from extragalactic supernovae, either from identified sources 
or from a diffuse flux due to unidentified past supernova explosions.

Observing neutrinos produced in the atmosphere as cosmic-ray secondaries
\cite{Aglietta:1988be,Hirata:1988uy,Hirata:1992ku,Becker-Szendy:1992hq,Daum:1994bf,Allison:1999ms,Ashie:2005ik} 
gave the first compelling evidence
for neutrino oscillation \cite{Fukuda:1998mi,Kajita:2006cy}, a process that unambiguously points to the existence of new physics. 
While today the puzzle of missing atmospheric neutrinos can be considered solved, 
there remain challenges related to the sub-dominant oscillation phenomena. In particular, precise measurements of
atmospheric neutrinos with high statistics and small systematic errors \cite{TabarellideFatis:2002ni}
would help in resolving ambiguities and degeneracies that hamper the interpretation 
of other experiments, as those planned for future long baseline neutrino oscillation measurements.

Another example of outstanding open questions is that of the knowledge of the interior of the Earth.  
It may look hard to believe, but we know much better what happens inside the Sun than inside our own planet. 
There are very few messengers that can provide information, while a mere theory is not sufficient for building a credible model for the Earth. However, there is a new unexploited window to the Earth's interior,
by observing neutrinos produced in the radioactive decays of heavy elements in the matter. Until now, only the KamLAND 
experiment  \cite{Araki:2005qa} has been able to study these so-called geo-neutrinos opening the way to a completely new
field of research.  The small event rate, however,  does not allow to draw significant conclusions. 

The fascinating physics phenomena outlined above, in addition to other important subjects that we will address in the following,
could be investigated by a new generation of multipurpose 
experiments based on improved detection techniques. 
The envisioned detectors must necessarily be very massive (and consequently large) 
due to the smallness of the cross-sections and to the low rate of signal events,
and able to provide very low experimental background. 
The required signal to noise ratio can only be achieved in underground laboratories suitably shielded against cosmic-rays
and environmental radioactivity.
We can identify three different and, to large extent, complementary technologies capable to meet the challenge, based
on large scale use of liquids for building large-size, volume-instrumented detectors

\begin{itemize}
\item Water Cherenkov.
As the cheapest available (active) target material, water is the only liquid that is realistic for extremely large detectors, 
up to several hundreds or thousands of ktons; \WC\ detectors have sufficiently good resolution in energy, 
position and angle. The technology is well proven, as previously used for the IMB, Kamiokande and Super-Kamiokande
experiments.

\item Liquid scintillator.
Experiments using a liquid scintillator as active target
provide high-energy resolution and offer low-energy threshold.  They are
particularly attractive for low energy particle detection, as for example solar
neutrinos and geo-neutrinos.  Also liquid scintillator detectors feature a well established technology, 
already successfully applied at relatively large scale to the Borexino
\cite{Back:2004zn} and KamLAND \cite{Araki:2004mb} experiments. 

\item Liquid Argon Time Projection Chambers (LAr TPC).
This detection technology has among the three the best performance in identifying the topology of
interactions and decays of particles, thanks to the bubble-chamber-like imaging performance. 
Liquid Argon TPCs are very versatile and work well with a wide particle energy range.
Experience on such detectors has been gained within the ICARUS project \cite{Amerio:2004ze,Arneodo:2001tx}.
\end{itemize}

Three experiments are proposed to employ the above detection techniques: MEMPHYS \cite{deBellefon:2006vq} for \WC,
LENA \cite{Oberauer:2005kw, Marrodan:2006} for liquid scintillator 
and GLACIER \cite{Rubbia:2004tz,Rubbia:2004yq,Ereditato:2004ru,Ereditato:2005ru,Ereditato:2005yx} for Liquid Argon. 
In this paper we report on the study of the physics potential of the experiments and identify features of complementarity 
amongst the three techniques. 

Needless to say, the availability of future neutrino beams from particle accelerators 
would provide an additional bonus to the above experiments.
Measuring oscillations with artificial neutrinos (of well known kinematical features)
with a sufficiently long baseline would allow to accurately determine the oscillation parameters
(in particular the mixing angle $\theta_{13}$ and the possible
CP violating phase in the mixing matrix). 
The envisaged detectors may then be used for observing neutrinos from the future Beta Beams and Super Beams
in the optimal energy range for each experiment. A common example 
is a low-energy Beta Beam from CERN to MEMPHYS at Frejus, 130 km away \cite{Campagne:2006yx}. 
High energy beams have been suggested \cite{Rubbia:2006pi}, 
favoring longer baselines of up to $O$(2000~km). 
The ultimate Neutrino Factory facility will require a magnetized detector to fully exploit the simultaneous availability of 
neutrinos and antineutrinos. This subject is however beyond the scope of the present study.

Finally, there is a possibility of (and the hope for) unexpected
discoveries. The history of physics has shown that 
several experiments have made their glory with discoveries in research fields that were outside the original goals of the experiments. 
Just to quote an example, we can mention the Kamiokande detector, mainly designed to search for proton decay
and actually contributing to the observation of atmospheric neutrino oscillations, to the clarification of the solar neutrino puzzle and 
to the first observation of supernova neutrinos \cite{Hirata:1987hu,Hirata:1988ad,Hirata:1989zj,Hirata:1988uy,
Fukuda:1998mi}.
All the three proposed experiments, thanks to their
outstanding boost in mass and performance, will certainly provide a significant potential for surprises and unexpected discoveries.

\section{Description of the three detectors}
\label{sec:Phys-detector}

The three detectors' basic parameters are listed in \refTab{tab:Phys-detector-summary}. 
All of them have active targets of tens to hundreds kton mass and are to be installed in underground laboratories to be protected against background induced by cosmic-rays. As already said,
the large size of the detectors is motivated by the extremely low cross-section of neutrinos and/or by the rareness of the 
interesting events searched for. Some details of the detectors are discussed in the following, while the matters related to the possible underground site are presented in Section~\ref{sec:Phys-Sites}.

\begin{table}
\caption{\label{tab:Phys-detector-summary}Basic parameters of the three detector (baseline) design.} 
\lineup
\begin{tabular}{@{}llll}
\br

                   &    GLACIER            &   LENA    &    MEMPHYS\\
\mr

\multicolumn{4}{@{}l}{Detector dimensions}          \\
type of cylinder              &  $1$ vert.    & $1$ horiz.    & $3\div5$ vert. \\
    diam. (m) & $\0 70$ & $\0 30$ & $\0 65$ \\	  
    length (m) & $\0 20$ & $100$ & $\0 65$ \\	  
typical mass (kton)   & $100$  &  $\0 50$  & $600\div800$\\
\mr
\multicolumn{4}{@{}l}{Active target and readout}          \\
	type of target  & liq. Argon      &liq. scintillator  & water \\
	                & (boiling)         &                      & (opt. 0.2\% GdCl$_3$) \\
readout type      & \parbox[t]{3cm}{
																		$e^-$ drift: 2 perp. views, $10^5$ channels, ampli. in gas phase;\\ 
																		Cher. light: $27\ 000$ 8" PMTs, $\sim 20\%$ coverage;\\
																		Scint. light: $1000$ 8" PMTs
																		}
                  & \parbox[t]{25mm}{$12\ 000$\\ 20" PMTs\\ $\gtrsim 30\%$ coverage} 
                  & \parbox[t]{25mm}{$81\ 000$\\ 12" PMTs\\$\sim 30\%$ coverage} \\
\br
\end{tabular}
\end{table}
%
\subsection{Liquid Argon TPC}

\begin{figure}
\begin{center}
\includegraphics[width=0.7\columnwidth]{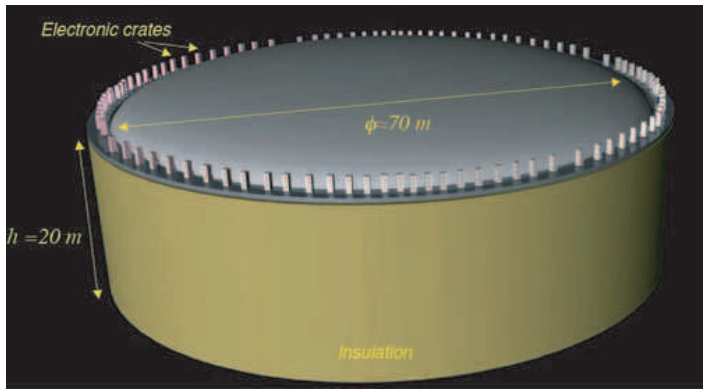}
\end{center}
\caption{\label{fig:Phys-GLACIERdetector} Artistic view of a 100~kton single-tank liquid Argon TPC detector. 
The electronic crates are located at the top of the dewar.}	
\end{figure}

GLACIER (Fig.~\ref{fig:Phys-GLACIERdetector}) is the foreseen extrapolation up to $100$~kton 
of the liquid Argon Time Projection Chamber technique. 
The detector can be mechanically subdivided into two parts,
the liquid Argon tank and the inner detector instrumentation.
For simplicity, we assume at this stage that the two aspects can be largely decoupled.
 
The basic idea behind this detector is to use a single 100~kton boiling liquid Argon cryogenic tank with
cooling directly performed with liquid Argon (self-refrigerating). Events are reconstructed in 3D by using the 
information provided by ionization in liquid. The imaging capabilities and the excellent space resolution
of the device make this detector an "electronic bubble chamber". 
The signal from scintillation and Cherenkov light readout complete the information contributing to the event reconstruction. 

As far as light collection is concerned one can profit from the ICARUS R\&D program that
has shown that it is possible to operate photomultipliers (PMTs) directly immersed in the liquid Argon \cite{Amerio:2004ze}. 
In order to be sensitive to DUV scintillation, PMTs are coated with a wavelength shifter (WLS), for instance
tetraphenyl-butadiene.
About 1000~immersed phototubes with WLS would
be used to identify the (isotropic and bright) scintillation light. To detect 
Cherenkov radiation about $27\ 000$~8''-phototubes without WLS would provide a 20\% coverage of the detector surface.
The latter PMTs should have single photon
counting capabilities in order to count the number of Cherenkov photons.

Charge amplification and an extreme  liquid purity against electronegative compounds
(although attainable by commercial purification systems) is needed to allow long drift distances of the ionization/imaging electrons
 ($\approx 20\rm\ m$). For this reason,
the detector will run in the so-called bi-phase mode. Namely, drifting electrons produced in the liquid phase 
are extracted into the gas phase with
the help of an electric field and amplified in order to compensate the charge loss due to 
attenuation along the drift path.
The final charge signal is then read out  by means of Large Electron Multiplier (LEM) devices, providing X-Y information. The Z coordinate
is given by the drift time measurement, proportional to the drift length.
A possible extension of the present detector design envisages the immersion of the sensitive volume in an external magnetic 
field \cite{Ereditato:2005yx}.
Existing experience from specialized Liquified Natural Gases (LNG) companies and studies conducted in collaboration with
Technodyne LtD UK,  have been ingredients for a first step in assessing the feasibility of the detector and of its operation
in an underground site.

\subsection{Liquid scintillator detector}

\begin{figure}
\begin{center}
\includegraphics[width=0.7\columnwidth]{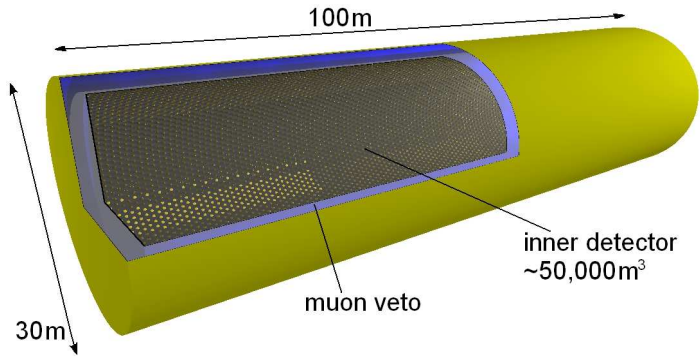}
\end{center}
\caption{\label{fig:Phys-LENAdetector}Schematic drawing of the LENA detector. Reprinted figure with the permission from \cite{Wurm:2007cy}.}	
\end{figure}

The LENA detector is cylindrical in shape with a length of about 100\,m and 30\,m diameter (\refFig{fig:Phys-LENAdetector}). 
The inner volume corresponding to a radius of 13\,m 
contains approximately $5 \times 10^4$\,m$^3$ of liquid scintillator.
The outer part of the volume is filled with water, acting as a
veto for identifying muons entering the detector from outside. 
Both the outer and the inner volume are enclosed in steel tanks
of 3 to 4\,cm wall thickness. For most purposes, a fiducial volume is defined by excluding
the volume corresponding to 1\,m distance to the inner tank walls. The fiducial volume so defined amounts
to 88\,$\%$ of the total detector volume.

The main axis of the cylinder is placed horizontally. A tunnel-shaped
cavern housing the detector is considered as realistically feasible for most of the envisioned detector locations. In
respect to accelerator physics, the axis could be oriented towards
the neutrino source in order to contain the full length of
muon and electron tracks produced in charged-current neutrino interactions in the liquid scintillator.

The baseline configuration for the light detection in the inner volume foresees 
$12\ 000$~PMTs of 20'' diameter mounted onto
the inner cylinder wall and covering about 30\,$\%$ of the surface. As
an option, light concentrators can be installed in front of the PMTs,
hence increasing the surface coverage $c$ to values larger than
50\,$\%$. Alternatively, $c=30\,\%$ can be reached by equipping
8'' PMTs with light concentrators, thereby reducing the cost when comparing to
the baseline configuration. Additional PMTs are supplied in the outer
veto to detect (and reject) the Cherenkov light from events due to incoming cosmic muons.
Possible candidates as liquid scintillator material are pure
phenyl-o-xylylethane (PXE), a mixture of 20\,$\%$ PXE and 80\,$\%$
Dodecane, and linear Alkylbenzene (LAB). All three liquids exhibit low 
toxicity and provide high flash and inflammation points.

\subsection{Water Cherenkov}

\begin{figure}
\begin{center}
\includegraphics[width=0.7\columnwidth]{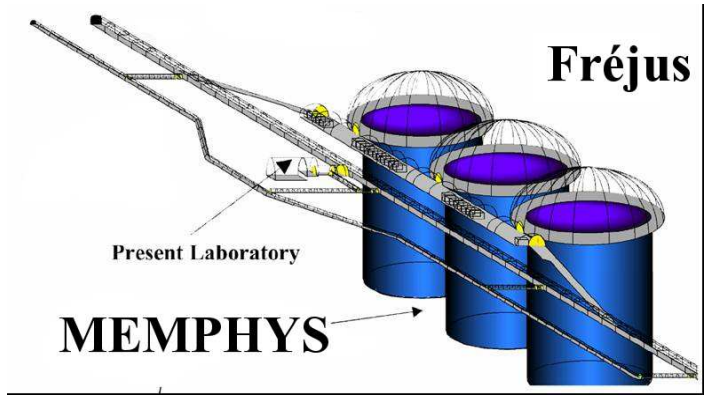}
\end{center}
\caption{\label{fig:Phys-MEMPHYSdetector}Layout of the MEMPHYS detector in the future Fréjus laboratory.}	
\end{figure}

The MEMPHYS detector (\refFig{fig:Phys-MEMPHYSdetector}) is an extrapolation of the  \WC\ Super-Kamiokande
detector to a mass as large as $730$~kton. 
The detector is composed of up to 5 shafts containing separate tanks. 
3 tanks are enough to total 440~kton fiducial mass. This is the configuration which is used hereafter. 
Each shaft has 65~m diameter and 65~m height representing an increase by a factor 8 with respect to Super-Kamiokande. 

The Cherenkov light rings produced by fast particles moving within the inner water volume are reconstructed by PMTs placed
on the inner tank wall.
The PMT housing surface starts at  2~m from the outer wall and is covered with about $81\ 000$ 12" PMTs to reach a 30\% surface coverage,
in or alternatively equivalent to a 40\% coverage with 20" PMTs. 
The fiducial volume is defined by an additional conservative guard of 2~m. 
The outer volume  between the PMT surface and the water vessel is instrumented with 8" PMTs. 
If not otherwise stated, the Super-Kamiokande analysis procedures for efficiency calculations, background reduction, etc.  are 
used in computing the physics potential of MEMPHYS. 
In USA and Japan, two analogous projects (UNO and Hyper-Kamiokande) have been proposed. 
These detectors are similar in many respects and the physics potential presented hereafter may well be transposed to them.
Specific characteristics that are not identical in the proposed projects are the distance from 
available or envisaged accelerators and nuclear reactors, sources of artificial neutrino fluxes, and the and the depth of the host laboratory. 

Currently, there is a very promising ongoing R\&D activity concerning 
the possibility of introducing Gadolinium salt (GdCl${}_3$) inside Super-Kamiokande. 
The physics goal is to decrease the background for many physics channels by detecting and tagging neutrons produced in 
the Inverse Beta Decay (IBD) interaction of $\bar{\nu}_e$ on free protons. 
For instance, 100~tons of GdCl${}_3$ in Super-Kamiokande would yield more then 90\% neutron captures on Gd  \cite{Beacom:2003nk}. 

\section{Underground sites}
\label{sec:Phys-Sites}

The proposed large detectors require underground laboratories of adequate size and depth, naturally protected against 
cosmic-rays that represent a potential source of background events mainly for non-accelerator experiments, that cannot exploit
the peculiar time stamp provided by the accelerator beam spill.

Additional characteristics of these sites contributing to their qualification as candidates for the proposed experiments 
are: the type and quality of the rock allowing the practical feasibility of large caverns at reasonable cost and within reasonable time, 
the distance from existing (or future) accelerators and nuclear reactors, the type and quality of the access, the geographical position, the environmental conditions, etc.

The presently identified worldwide candidate sites are located in three geographical regions: North-America, far-east Asia 
and Europe. In this paper we consider the European region, where, at this stage, the following sites 
are assumed as candidates: Boulby (UK), Canfranc (Spain), Fréjus (France/Italy), Gran Sasso (Italy), 
Pyhäsalmi (Finland) and Sieroszewice (Poland). 
Most of the sites are existing national or international underground laboratories with associated infrastructure
and experimental halls already used for experiments.
The basic features of the sites are presented on \refTab{tab:Phys-site-parameters}. 
For the Gran Sasso Laboratory a  possible new (additional) site 
is envisaged to be located 10 km away from the present underground laboratory, 
outside the protected area of the neighboring Gran Sasso National Park.
The possibility of under-water solutions, such as for instance Pylos for the LENA project, is not taken into account here.
The identification and measurement of the different background components in the candidate sites (muons, fast neutrons 
from muon interactions, slow neutrons from nuclear reactions in the rock, gammas, electrons/positrons and alphas from 
radioactive decays,\dots) is underway, mainly in the context of the ILIAS European (JRA) Network ($http://ilias.in2p3.fr/$). 

None of the existing sites has yet a sufficiently large cavity able to accommodate the foreseen detectors.
For two of the sites (Fréjus and Pyhäsalmi) a preliminary feasibility study for large excavation at deep depth 
has already been performed. For the Fréjus site the main conclusion drawn from simulations constrained by a series 
of rock parameter measurements made during the Fréjus road tunnel excavation is that the "shaft shape" is strongly preferred 
compared to the "tunnel shape", as long as large cavities are required. As mentioned above, 
several (up to 5) of such shaft cavities with a diameter of about 65~m 
(for a corresponding volume of $250\ 000$~m${}^3$) each, seem feasible in the region around the middle of the Fréjus tunnel, at a depth of 4800~m.w.e.
For the Pyhäsalmi site, the preliminary study has been performed for two main cavities with tunnel shape and 
dimensions of $(20 \times 20 \times 120)$~m${}^3$ and $(20 \times 20 \times 50)$~m${}^3$, respectively, 
and for one shaft-shaped cavity with 25~m in diameter and 25~m in height, all at a depth of about 1430~m of rock (4000~m.w.e.). 

\begin{sidewaystable}
\caption{\label{tab:Phys-site-parameters} 
Summary of characteristics of some underground sites envisioned for the proposed detectors.} 
\begin{tabular}{@{}lllllll} 
\br 
Site &     Boulby      &       Canfranc          &      Fréjus     &  Gran Sasso   &   Pyh\"asalmi  & Sieroszowice\\ 
\mr 
Location &    UK       &      Spain           &   Italy-France border &      Italy  &      Finland  &  Poland     \\ 
Dist. from CERN (km)&  1050  &  630          &          130       &     730        &     2300     &    950       \\ 
Type of access&  Mine  &  Somport tunnel     &  Fréjus tunnel     & Highway\\ tunnel &  Mine        &   Shaft      \\ 
Vert. depth (m.w.e)&  2800 & 2450           &    4800           &   3700       &  4000         &  2200       \\ 
Type of rock& salt     &   hard rock          &  hard rock         & hard rock     & hard rock      &  salt \& rock \\ 
 Type of cavity&       &                       &   shafts          &               &   tunnel       &    shafts    \\ 
Size of cavity &       &                       & $\Phi = 65~\mathrm{m}$ &          & $(20\times20\times 120)\mathrm{m^3}$          & $\Phi = 74~\mathrm{m}$             \\ 
         &             &                         & $H=80~\mathrm{m}$ &             &                & $H=37~\mathrm{m}$ \\ 
$\mu$ Flux (m$^{-2}$day$^{-1}$)&  34 & 406 &             4         &    24         &      9          &  not available            \\ 
\br
\end{tabular} 
%
\end{sidewaystable}
%



\section{Matter instability: sensitivity to proton decay}

\begin{figure}
\begin{center}
\includegraphics[width=0.7\columnwidth]{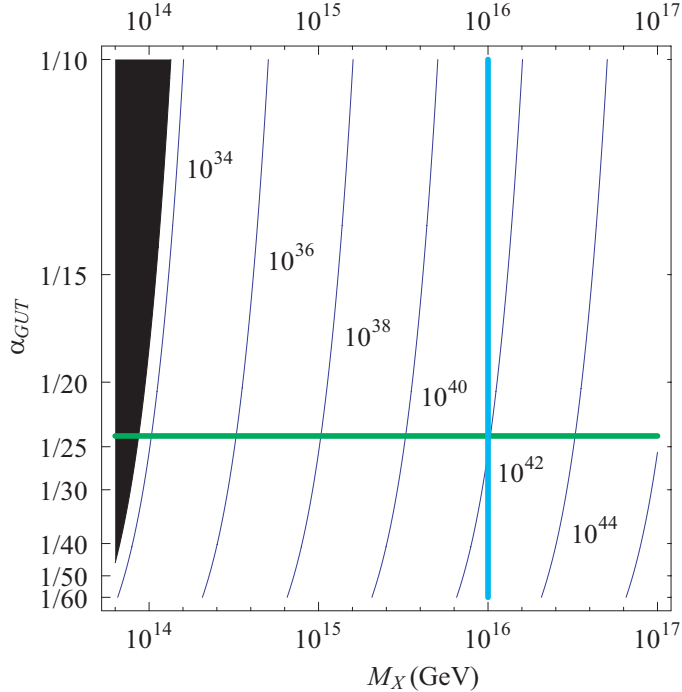}
\end{center}
\caption{\label{fig:Phys-PDK-Majorana} Isoplot for the upper bounds on the total
proton lifetime in years in the Majorana neutrino case in the
$M_X$--$\alpha_{GUT}$ plane. The value of the unifying coupling
constant is varied from $1/60$ to $1/10$. The conventional values
for $M_X$ and $\alpha_{GUT}$ in SUSY GUTs are marked with thick
lines. The experimentally excluded region is given in black. Reprinted figure with permission from~\cite{Dorsner:2004xa}.}
\end{figure}

For all relevant aspects of the proton stability in Grand Unified Theories, 
in strings and in branes we refer to~\cite{Nath:2006ut}.   
Since proton decay is the most dramatic prediction coming
from theories of the unification of fundamental interactions, there is a realistic hope to be able to test these scenarios with next 
generation experiments exploiting the above mentioned large mass, underground detectors.
For this reason, the knowledge of a theoretical upper bound on the lifetime of the proton is very 
helpful in assessing the potential of future experiments.   
Recently, a model-independent upper bound on the proton decay lifetime has 
been worked out~\cite{Dorsner:2004xa}

\begin{equation}
\fl
	\tau_p^{upper} = 	
		\left\{\begin{array}{lr}
	6.0 \times 10^{39} & (\mathrm{Majorana}) 
         \\ 
         2.8 \times 10^{37}  & (\mathrm{Dirac})
	\end{array}\right\}
                 \times 
	 \frac{\left(M_X/10^{16}GeV\right)^4}{\alpha_{GUT}^2} \times \left( \frac{0.003GeV^3}{\alpha} \right)^2 \ \mathrm{years}         
\end{equation}
where $M_X$ is the mass of the superheavy gauge bosons mediating 
proton decay, the parameter $\alpha_{GUT}= g_{GUT}^2 / 4 \pi$, 
with $g_{GUT}$ the gauge coupling at the grand unified scale 
and $\alpha$ is the relevant matrix element.
\refFig{fig:Phys-PDK-Majorana} shows the present parameter space allowed by experiments
in the case of Majorana neutrinos.

Most of the models (Super-symmetric or non Super-symmetric) predict a proton lifetime $\tau_p$ below 
those upper bounds ($10^{33-37}$~years). This is  particularly interesting since this falls within the possible 
range of the proposed experiments. 
In order to have a better idea of the proton decay predictions, we list 
the results from different models in \refTab{tab:Phys-PDK-Models}.

No specific simulations for MEMPHYS have been carried out yet. Therefore,
here we rely on the studies done for the similar UNO detector, adapting the results to MEMPHYS, which, however, features
an overall better PMT coverage. 

\begin{table}
\caption{\label{tab:Phys-PDK-Models}
Summary of several predictions for the proton partial lifetimes (years). References for the 
different models are: (1) \cite{Georgi:1974sy}, (2) \cite{Dorsner:2005fq,Dorsner:2005ii}, (3) \cite{Lee:1994vp},  
(4)  \cite{Murayama:2001ur,Bajc:2002bv,Bajc:2002pg,Emmanuel-Costa:2003pu}, 
(5) \cite{Babu:1992ia,Aulakh:2003kg,Fukuyama:2004pb,Goh:2003nv}, 
(6) \cite{Friedmann:2002ty},
(7) \cite{Bajc:2006ia},
(8) \cite{Perez:2007rm}.}
\begin{tabular}{@{}llll} \br
Model       &   Decay modes     &  Prediction   &  References \\ \mr
Georgi-Glashow model & - &  ruled out      &        (1)        \\ 
\parbox{4cm}{\center{Minimal realistic\\ non-SUSY $SU(5)$}} & all channels & $\tau_p^{upper} = 1.4 \times 10^{36}$ & (2)
\\[6mm]
Two Step Non-SUSY $SO(10)$ &  $p \to e^+ \pi^0$ &  $\approx 10^{33-38}$ & (3)  \\[5mm] 
Minimal SUSY $SU(5)$   &   $p \to  \bar{\nu}K^+$  &  $\approx 10^{32-34}$  & (4)
\\ 
\\[-5mm]
\parbox{4cm}{\center{SUSY $SO(10)$ \\ with $10_H$, and $126_H$}} & $p \to \bar{\nu} K^+$ & $\approx 10^{33-36}$ &  (5)  
\\[6mm]
M-Theory($G_2$)   & $p \to e^+\pi^0$    &  $\approx 10^{33-37}$    & (6)  \\[4mm]
 $SU(5)$ with $24_F$  & $p \to \pi^0 e^+ $ & $\approx 10^{35-36}$   & (7)\\[4mm]
 Renormalizable Adjoint $SU(5)$ & $p \to \pi^0 e^+ $ & $\approx 10^{35-36}$   & (8)\\
\br
		\end{tabular}
\end{table}

In order to assess the physics potential of a large liquid Argon Time Projection Chambers such as GLACIER, 
a detailed simulation of signal efficiency and
background sources, including atmospheric neutrinos and cosmogenic
backgrounds was carried out \cite{Bueno:2007um}. Liquid Argon TPCs,
offering high space granularity and energy resolution, low-energy detection threshold,
and excellent background discrimination, should  
yield  large signal over background ratio for many of the possible proton
decay modes, hence allowing reaching partial lifetime sensitivities in
the range of $10^{34}-10^{35}$~years for exposures up to 1000~kton year.
This can often be accomplished in quasi background-free conditions optimal for discoveries 
at the few events level, corresponding
to atmospheric neutrino background rejections of the order of $10^5$.

Multi-prong decay modes like $p\rightarrow \mu^- \pi^+ K^+$
or $p\rightarrow e^+\pi^+\pi^-$ and channels involving kaons like
$p\rightarrow K^+\bar\nu$, $p\rightarrow e^+K^0$ and $p\rightarrow \mu^+K^0$
are particularly appealing, since liquid Argon imaging
provides typically one order of magnitude efficiency increase for similar
or better background conditions, compared to water Cherenkov detectors.
Up to a factor of two improvement in efficiency is expected for modes like $p\rightarrow e^+\gamma$
and $p\rightarrow \mu^+\gamma$, thanks to the clean photon identification
and separation from $\pi^0$. Channels such as $p\rightarrow e^+\pi^0$ and $p\rightarrow \mu^+\pi^0$,
dominated by intrinsic nuclear effects,
yield similar performance as water Cherenkov detectors. 

An important feature of GLACIER is that thanks to the self-shielding 
and 3D-imaging properties, the above expected performance
remains valid even at shallow depths, where cosmogenic background sources are important.
The possibility of using a very large-area, annular, muon-veto active shielding, to
further suppress cosmogenic backgrounds at shallow depths is also a very promising 
option to complement the GLACIER detector.

In order to quantitatively estimate the potential of the LENA detector
in measuring proton lifetime, a Monte Carlo simulation for the
decay channel $p\to K^{+}\overline{{\nu}}$ has been performed. For
this purpose, the GEANT4 simulation toolkit \cite{Agostinelli:2002hh} has been
used, including optical processes as scintillation, Cherenkov light
production, Rayleigh scattering and light absorption. From these simulations one obtains
a light yield  of $\sim 110$~p.e./MeV for an event in the
center of the detector. In  addition, the semi-empirical Birk's formula
has been introduced into the code in order to take into account the so-called quenching effects.

\begin{figure}
\begin{center}
\includegraphics[width=0.7\columnwidth]{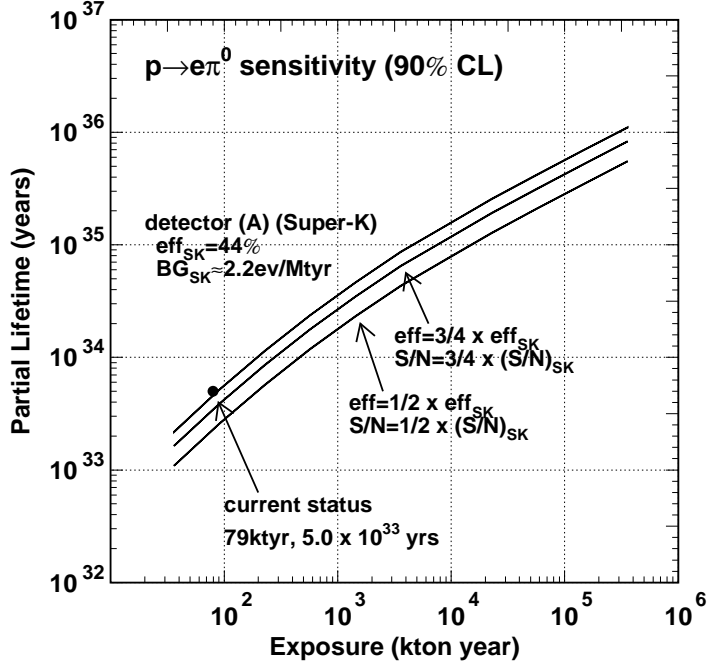}
\end{center}
\caption{\label{fig:pdk1} Sensitivity to the $e^+\pi^0$ proton decay mode
compiled by the UNO collaboration. MEMPHYS corresponds to case (A). Reprinted figure with permission from~\cite{Jung:1999jq}.}
\end{figure}

Following studies performed for the UNO detector, the detection efficiency for $p \rightarrow e^+\pi^0$
is $43\%$ for a 20" PMT coverage of 40\% or its equivalent, as envisioned for
MEMPHYS. The corresponding estimated
atmospheric neutrino induced background is at the level of $2.25$~events/Mton year. 
From these efficiencies and background levels,
proton decay sensitivity as a function of detector exposure can be
estimated. A $10^{35}$ years partial
lifetime ($\tau_p/B$) could be reached at the 90\% C.L. for a 5~Mton year exposure (10~years) with MEMPHYS
(similar to case A in \refFig{fig:pdk1} compiled by the UNO collaboration \cite{Jung:1999jq}). Beyond that exposure, tighter cuts may be envisaged to further reduce the atmospheric neutrino background to $0.15$~events/Mton year, by selecting quasi exclusively the free proton decays.

\begin{figure}
\begin{center}
\includegraphics[width=0.7\columnwidth]{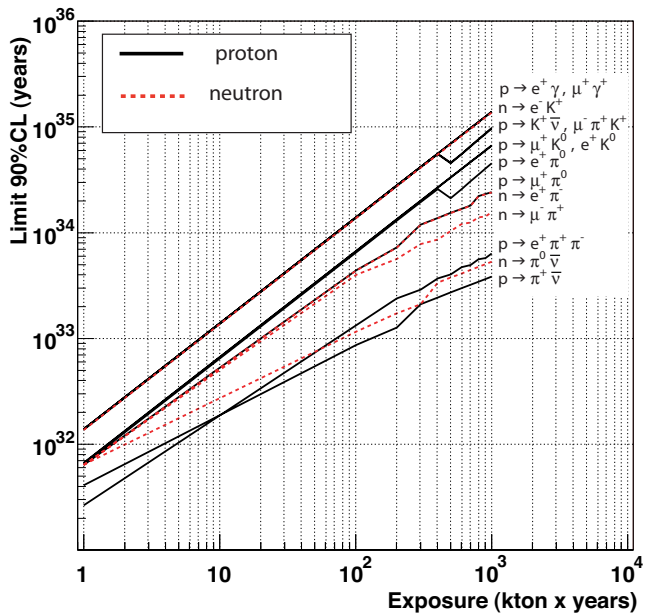}
\end{center}
\caption{\label{fig:GLACIERpdk} Expected proton decay lifetime limits ($\tau / B$ at 90\% C.L.) 
as a function of exposure for GLACIER. Only atmospheric neutrino background 
has been taken into account. Reprinted figure with permission from~\cite{Bueno:2007um}.}
\end{figure}

The positron and the two photons issued from the $\pi^0$ gives clear events 
in the GLACIER detector. The $\pi^0$ is absorbed by the nucleus
in $45\%$ of the cases. Assuming a perfect particle and track identification, 
one may expect a $45\%$ efficiency and a background level of $1$~event/Mton year. 
For a 1~Mton year (10~years) exposure with GLACIER one 
reaches $\tau_p/B > 0.4 \times 10^{35}$~years at the 90$\%$ C.L. (Fig.~\ref{fig:GLACIERpdk}). 

In a liquid scintillator detector such as LENA the decay $p\to e^{+}\pi^{0}$ would
produce a $938$~MeV signal coming from the $e^{+}$ and the $\pi^{0}$
shower. Only atmospheric neutrinos are expected to cause background
events in this energy range. Using the fact that showers from both
$e^{+}$ and $\pi^{0}$ propagate 4~m in opposite directions
before being stopped, atmospheric neutrino background can be
reduced. Applying this method, the current limit for this channel
($\tau_p/B=5.4~10^{33}$~years \cite{Nakaya:2005nk}) could be improved.
In LENA, proton decay events via the mode $p\to K^{+}\overline{{\nu}}$
have a very clear signature. The kaon causes a prompt monoenergetic
signal of 105~MeV together with a larger delayed signal from its decay.
The kaon has a lifetime of 12.8~ns and two main decay channels: with a
probability of 63.43~$\%$ it decays via $K^{+}\to\mu^{+}{\nu_{\mu}}$
and with 21.13\%, via \mbox{$K^{+}\to\pi^{+} \pi^{0}$}.

Simulations of proton decay events and atmospheric neutrino background
have  been performed and a pulse shape analysis has been applied.
From this analysis an efficiency of 65\% for
the detection of a proton decay has been determined and a
background  suppression of $\sim2 \times10^{4}$ has  been
achieved \cite{Undagoitia:1-2uu}. A detail study of background implying pion and
kaon production in atmospheric  neutrino reactions has been performed
leading to a background rate of $0.064~\mathrm{year}^{-1}$ due to the reaction
${\nu}_{\mu}+p\to \mu^{-}+K^{+}+p$.

For the current proton lifetime limit for the channel considered
($\tau_p/B=2.3 \times 10^{33}$~year) \cite{Kobayashi:2005pe}, about 40.7 proton decay
events would be observed in LENA after ten years
with less than 1 background event. If no signal is seen in the detector
within ten years, the lower limit for the lifetime of the proton
will be set at $\tau_p/B>4~\times10^{34}$~years at the $90\%$~C.L. 

For GLACIER, the latter is a quite clean
channel due to the presence of a strange meson and no other particles in
the final state. Using $dE/dx$ versus range as the discriminating variable 
in a Neural Network algorithm, less than $1\%$ of the kaons are mis-identified as protons. 
For this channel, the selection efficiency is high ($97\%$) 
for an atmospheric neutrino background $< 1$~event/Mton year. 
In case of absence of signal and for a detector location at a depth of 
1~km.w.e., one expects for 1~Mton~year (10~years) exposure one background event due to cosmogenic sources. This translates into a limit 
$\tau_p/B > 0.6 \times 10^{35}$~years at 90\% C.L. This result remains 
valid even at shallow depths where
cosmogenic background sources are a very important limiting factor for proton 
decay searches. 
For example, the study done in \cite{Bueno:2007um} shows that 
a three-plane active veto at a shallow
depth of about 200~m rock overburden under a hill yields
similar sensitivity for $p\rightarrow K^+\bar\nu$ as a 3000~m.w.e. deep detector.

\begin{figure}
\begin{center}
\includegraphics[width=0.7\columnwidth]{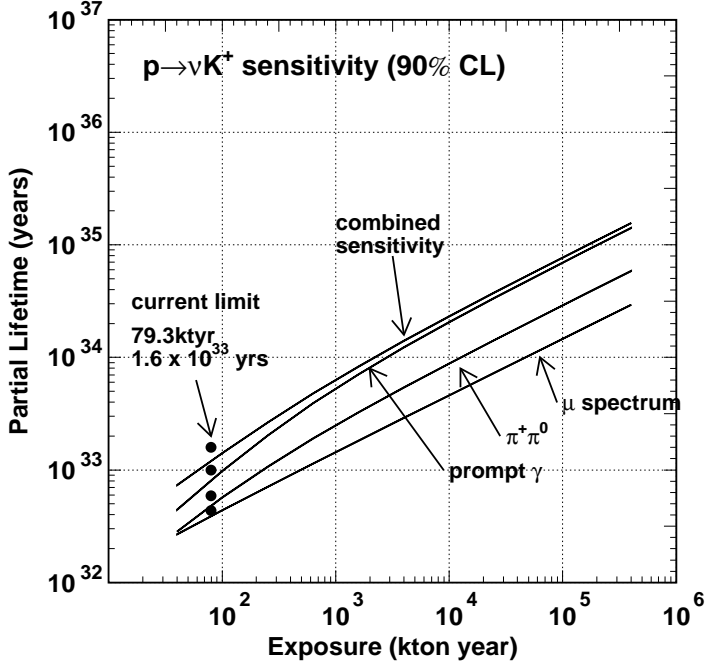}
\end{center}
\caption{\label{fig:pdk9_jbz}
Expected sensitivity to the $\nu K^+$ proton decay mode as a function of
exposure compiled by the UNO collaboration which may be applied for the MEMPHYS detector (see text for details). Reprinted figure with permission from~\cite{Jung:1999jq}.}
\end{figure}

For MEMPHYS one should rely on the detection of the decay products of the $K^+$
since its momentum ($360$~MeV) is below the water Cherenkov threshold of $570$~MeV: a 256~MeV/c muon and its 
decay electron (type I) or a 205~MeV/c $\pi^+$ and $\pi^0$
(type II), with the possibility of a delayed (12~ns) coincidence
with the 6~MeV ${}^{15}\mathrm{N}$ de-excitation prompt $\gamma$ (Type III).
Using the known imaging and timing performance of Super-Kamiokande, the efficiency for the reconstruction of
$p \rightarrow \overline{\nu}K^+$ is 33\% (I), 6.8\% (II)
and 8.8\% (III), and the background is 2100, 22 and 6 events/Mton year, respectively. For the
prompt $\gamma$ method, the background is dominated by
miss-reconstruction. As stated by the UNO Collaboration \cite{Jung:1999jq}, there are good 
reasons to believe that this background can be lowered by at least a factor of two, corresponding
to the atmospheric neutrino interaction $\nu p \rightarrow \nu
\Lambda K^+$. In these conditions, and taking into account the Super-Kamiokande performance,
a 5~Mton year exposure for MEMPHYS would allow reaching $\tau_p/B > 2 \times 10^{34}$~years (\refFig{fig:pdk9_jbz}).

A preliminary comparison between the performance of  three detectors has been carried out 
(Tab.~\ref{tab:Phys-PDK-Summary}). 
For the $e^+ \pi^0$ channel, the Cherenkov detector gets a better limit due to the
higher mass. However, it should be noted that GLACIER, although five times smaller 
in mass than MEMPHYS,  can reach a limit that is only a factor two smaller. 
Liquid Argon TPCs and liquid scintillator detectors obtain better results for the
$\bar{\nu} K^+$ channel, due to their higher detection efficiency.
The techniques look therefore quite complementary. 
We have also seen that GLACIER does not necessarily requires very deep underground 
laboratories, like those currently existing or future planned sites, in order to perform high 
sensitivity nucleon decay searches.

\begin{table}
\caption{\label{tab:Phys-PDK-Summary}Summary of the $e^+\pi^0$ and $\bar{\nu}K^+$ decay
discovery potential for the three detectors. 
The $e^+\pi^0$ channel is not yet simulated for LENA.}
\begin{indented}
\item[]\begin{tabular}{@{}llll}\br
						& GLACIER             &      LENA              &  MEMPHYS \\ \mr
$e^+\pi^0$	&                     &                        &          \\
$\epsilon (\%)
/ \mathrm{Bkgd (Mton~year)}$ & $45/1$  &         -               &   $43/2.25$ \\
$\tau_p/B$ (90\% C.L., 10~years) & 	$0.4\times 10^{35}$ & -           &  $1.0\times 10^{35}$ \\ \mr

$\bar{\nu}K^+$                    &                         &              \\
$\epsilon (\%)
/ \mathrm{Bkgd (Mton~ year)}$ & $97/1$  &         $65/1$               &   $8.8/3$ \\
$\tau_p/B$ (90\% C.L., 10~years) & 	$0.6\times 10^{35}$ & $0.4\times 10^{35}$            &  $0.2\times 10^{35}$ \\
 \br
\end{tabular}
\end{indented}
\end{table}

\section{Supernova neutrinos}
\label{sec:SN}

The detection of supernova (SN) neutrinos represents one of the next
frontiers of neutrino physics and astrophysics. It will provide invaluable
information on the astrophysics of the core-collapse explosion
phenomenon and on the neutrino mixing parameters. In particular,
neutrino flavor transitions in the SN envelope might be sensitive
to the value of $\theta_{13}$ and to the type of mass hierarchy. 
These two main issues are discussed in detail in the following Sections.

\subsection{SN neutrino emission, oscillation and detection}

A core-collapse supernova marks the evolutionary end of a massive star
($M\gtrsim 8\,M_\odot$) which becomes inevitably unstable at the end
of its life. The star collapses and ejects its outer mantle in a shock-wave
driven explosion.  The collapse to a neutron star ($M \simeq M_\odot
$, $R\simeq 10$~km) liberates a gravitational binding energy of
$\approx 3 \times10^{53}~{\rm erg} $, 99\% of which is transferred to 
(anti) neutrinos of all the flavors and only 1\% to the
kinetic energy of the explosion. Therefore, a core-collapse SN
represents one of the most powerful sources of (anti) neutrinos in the Universe.
In general, numerical simulations of SN explosions provide the
original neutrino spectra in energy and time $F^0_{\nu}$. Such initial
distributions are in general modified by flavor transitions in the SN
envelope, in vacuum (and eventually in Earth matter): $F^0_\nu {\longrightarrow} F_\nu$
and must be convoluted with the differential interaction cross-section
$\sigma_e$ for electron or positron production, as well as with the
detector resolution function $R_e$ and the efficiency $\varepsilon$,
in order to finally get observable event rates $N_e = F_\nu \otimes \sigma_e \otimes R_e \otimes \varepsilon $.

Regarding the initial neutrino distributions $F^0_{\nu}$, a SN
collapsing core is roughly a black-body source of thermal neutrinos,
emitted on a timescale of $\sim 10$~s.  Energy spectra parametrizations
are typically cast in the form of quasi-thermal distributions, with
typical average energies: $ \langle E_{\nu_e} \rangle= 9-12$~MeV,
$\langle E_{\bar{\nu}_e} \rangle= 14-17$~MeV, $\langle E_{\nu_x}
\rangle= 18-22$~MeV, where $\nu_x$ indicates any non-electron flavor.

The oscillated neutrino fluxes arriving on Earth may be
written in terms of the energy-dependent  survival probability
 $p$ ($\bar{p}$) for neutrinos (antineutrinos) as \cite{Dighe:1999bi}

\begin{eqnarray}
F_{\nu_e} & = & p F_{\nu_e}^0 + (1-p) F_{\nu_x}^0  \nonumber \\ 
F_{\bar\nu_e} & =  &\bar{p} F_{\bar\nu_e}^0 + (1-\bar{p}) F_{\nu_x}^0 \label{eqfluxes1-3} \\
4 F_{\nu_x} & = & (1-p) F_{\nu_e}^0 + (1-\bar{p}) F_{\bar\nu_e}^0 +
(2 + p + \bar{p}) F_{\nu_x}^0 \nonumber
\end{eqnarray}

where $\nu_x$ stands for either $\nu_\mu$ or $\nu_\tau$.  The
probabilities $p$ and $\bar{p}$ crucially depend on the neutrino mass
hierarchy and on the unknown value of the mixing angle $\theta_{13}$
as shown in \refTab{tab:Phys-SN-Flux}.

\begin{table}
		\caption{\label{tab:Phys-SN-Flux}Values of the $p$ and $\bar{p}$ parameters used in
 Eq.~\ref{eqfluxes1-3} in different scenario of mass hierarchy and  $\sin^2 \theta_{13}$.}
\begin{indented}
\item[]\begin{tabular}{@{}llll} \br
		Mass Hierarchy        & $\sin^2\theta_{13}$ & $p$     & $\bar{p}$ \\ \mr
		Normal                & $\gtrsim 10^{-3}$              & 0        & $\cos^2 \theta_{12}$ \\ 
		Inverted		          & $\gtrsim 10^{-3}$              & $\sin^2 \theta_{12}$ & 0 \\
		Any                   &  $\lesssim 10^{-5}$             & $\sin^2 \theta_{12}$ & $\cos^2 \theta_{12}$ \\
\br
		\end{tabular}
\end{indented}
\end{table}
Galactic core-collapse supernovae are rare, perhaps a few per century. 
Up to now, SN neutrinos have been detected only once
during the SN~1987A explosion in the Large Magellanic Cloud in 1987 ($d=50$~kpc).
Due to the relatively small masses of the detectors operational at that time,  only few events were detected:
11 in Kamiokande \cite{Hirata:1987hu,Hirata:1988ad} and 8 in IMB \cite{Aglietta:1987we,Bionta:1987qt}. 
The  three proposed large-volume neutrino observatories can guarantee continuous exposure for 
several decades, so that  a high-statistics SN neutrino signal could be eventually observed.
The expected number of events for GLACIER, LENA and MEMPHYS
are reported in \refTab{tab:Phys-SN-DetectorRates} for a typical galactic SN distance
of $10$~kpc. 
The total number of events is shown in the upper panel, while the lower part refers to the $\nu_e$ signal detected
during the prompt neutronization burst, with a duration of $\sim 25$~ms, just after the core bounce. 

\begin{table}
		\caption{\label{tab:Phys-SN-DetectorRates} Summary of the expected neutrino interaction 
rates in the different detectors for a typical SN.
The following notations have been used: CC, NC, IBD, $e$ES and pES stand for Charged Current, Neutral Current, Inverse Beta Decay, 
electron and proton Elastic Scattering, respectively. The final state nuclei are generally unstable and decay either 
radiatively (notation ${}^*$), or by $\beta^-/\beta^+$ weak interaction (notation ${}^{-,+}$).
The rates of the different reaction channels are listed, and for LENA they have been obtained by scaling
the predicted rates from \cite{Cadonati:2000kq, Beacom:2002hs}.}
%
\lineup
\begin{tabular}{@{}llllll} \br
		\multicolumn{2}{@{}c}{MEMPHYS} & \multicolumn{2}{c}{LENA} & \multicolumn{2}{c}{GLACIER} \\ \ns
		Interaction    & Rates  & Interaction    & Rates  & Interaction    & Rates  \\ \mr
		$\bar{\nu}_e$ IBD & $2 \times 10^{5}$ &
		$\bar{\nu}_e$ IBD & $9.0 \times 10^{3}$ &
		$\nu_e^{CC}({}^{40}Ar,{}^{40}K^*)$ & $2.5 \times 10^{4}$ \\
		$\nunubar{e}{}^{CC} ({}^{16}O,X) $ & $1\times10^{4}$ &
		$\nu_x$ pES  & $7.0 \times 10^{3}$ &
		$\nu_x^{NC}({}^{40}Ar^{*})$ & $3.0 \times 10^{4}$ \\		 
		$\nu_x$ $e$ES  & $1\times10^{3}$ &
		$\nu_x^{NC} ({}^{12}C^{*})$ & $3.0 \times 10^{3}$ &
		$\nu_x$ $e$ES & $1.0\times10^{3}$ \\
		& & 
		$\nu_x$ $e$ES & $6.0\times 10^2$ &
		$\bar{\nu}_e^{CC}({}^{40}Ar,{}^{40}Cl^*)$ & $5.4 \times 10^2$ \\
	  &		&
		$\bar{\nu}_e^{CC} ({}^{12}C,{}^{12}B^{+})$ & $5.0\times10^2$ & &\\
		& &
		$\nu_e^{CC} ({}^{12}C,{}^{12}N^{-})$ & $8.5 \times 10^1$  & & \\
		\mr
		\multicolumn{6}{@{}l}{Neutronization Burst rates}\\
		  MEMPHYS & $\0 60$ & ${\nu}_e$ eES & & & \\
		    LENA & 
		    $\0 70$ & $\nu_e$ eES/pES & &  & \\
		    
		    GLACIER & $380$ & $\nu_x^{NC}({}^{40}Ar^{*})$ & & & \\
		\br
		\end{tabular}
\end{table}

The $\bar{\nu}_e$ detection by IBD 
is the golden channel for MEMPHYS and LENA. In addition, the electron neutrino signal can be detected by LENA
thanks to the interaction on $^{12}$C.  The three charged-current reactions would provide 
information on $\nu_e$ and $\bar{\nu}_{e}$ fluxes and spectra while the three neutral-current processes,
sensitive to all neutrino flavours, would give information on the total flux.
GLACIER has also the opportunity to detect $\nu_e$ by charged-current
interactions on ${}^{40}\rm{Ar}$ with a very low energy threshold. 
The detection complementarity between $\nu_e$ and $\bar{\nu}_e$ is of
great interest and would assure a unique way of probing the SN explosion
mechanism as well as assessing intrinsic neutrino properties.  Moreover, the
huge statistics would allow spectral studies in time and in energy domain.

We wish to stress that it will be difficult to establish SN neutrino
oscillation effects solely on the basis of a $\bar\nu_e$ or $\nu_e$
spectral hardening, relative to theoretical expectations. Therefore, in the recent literature the importance of
model-independent signatures has been emphasized. Here we focus
mainly on signatures associated to the prompt $\nu_e$
neutronization burst, the shock-wave propagation and the Earth matter crossing.

The analysis of the time structure of the SN signal during the first few tens of milliseconds
after the core bounce can provide a clean indication if the full $\nu_e$ burst is present or
absent, and therefore allows distinguishing between different mixing scenarios, as indicated by the
third column of \refTab{tab:Phys-SN-SummaryOscNeut}. For example, if the mass
ordering is normal and $\theta_{13}$ is large, the $\nu_e$ burst
will fully oscillate into $\nu_x$.  If $\theta_{13}$ turns out to be relatively large
one could be able to distinguish between normal and inverted neutrino mass hierarchy.  

As discussed above, MEMPHYS is mostly sensitive to the IBD, although
the $\nu_e$ channel can be measured by the elastic scattering reaction
$\nu_x+e^-\to e^-+\nu_x$ \cite{Kachelriess:2004ds}. Of course, the
identification of the neutronization burst is the 
cleanest with a detector exploiting the charged-current absorption of $\nu_e$ neutrinos, such as
GLACIER.  Using its unique features of measuring $\nu_e$ CC (Charged Current) events it is
possible to probe oscillation physics during the early stage of the SN explosion, while with NC (Neutral Current) events one can 
decouple the SN
mechanism from the oscillation physics \cite{Gil-Botella:2004bv,Gil-Botella:2003sz}. 

A few seconds after core bounce, the SN shock wave will pass the density region in the stellar envelope relevant for oscillation matter
effects, causing a transient modification of the survival probability and thus a time-dependent signature in the neutrino signal
\cite{Schirato:2002tg,Fogli:2003dw}.  This would produce a characteristic
dip when the shock wave passes \cite{Fogli:2004ff}, or a double-dip if a reverse shock occurs \cite{Tomas:2004gr}. The
detectability of such a signature has been studied in a large \WC\
detector like MEMPHYS by the IBD \cite{Fogli:2004ff}, and in a
liquid Argon detector like GLACIER by Argon CC interactions
\cite{Barger:2005it}. The shock wave effects would certainly be
visible also in a large volume scintillator such as LENA. Such observations
would test our theoretical understanding of the core-collapse SN phenomenon, in addition to identifying the actual 
neutrino mixing scenario.
 
Nevertheless, the supernova matter profile need not be smooth. Behind the
shock-wave, convection and turbulence can cause significant stochastic density
fluctuations which tend to cast a shadow by making other features, such as the shock front, 
unobservable in the density range covered by the turbulence \cite{Fogli:2006xy,Friedland:2006ta}. The quantitative
relevance of this effect remains to be understood. 

A unambiguous indication of oscillation effects would be the energy-dependent modulation of the survival probability 
$ p(E)$ caused by Earth matter effects \cite{Lunardini:2001pb}. 
These effects can be revealed by peculiar wiggles in the energy spectra, due
to neutrino oscillations in Earth crossing.
In this respect, LENA benefits from a better energy resolution than MEMPHYS, which may be partially compensated by 
10 times more statistics
\cite{Dighe:2003jg}.  The Earth effect would show up in the $\bar\nu_e$ channel for the normal mass hierarchy, assuming 
that $\theta_{13}$ is large (\refTab{tab:Phys-SN-SummaryOscNeut}). Another possibility to establish the presence of Earth 
effects is to use the signal from two detectors if one of them sees the SN shadowed by the
Earth and the other not. A comparison between the signal normalization in the two detectors might reveal Earth 
effects \cite{Dighe:2003be}.
The probability for observing a Galactic SN shadowed by the Earth as
a function of the detector's geographic latitude depends only mildly
on details of the Galactic SN distribution \cite{Mirizzi:2006xx}. A location at the
North Pole would be optimal with a shadowing probability of about
60\%, but a far-northern location such as Pyh\"asalmi in Finland, the
proposed site for LENA, is almost equivalent (58\%). One particular
scenario consists of a large-volume scintillator detector located in
Pyh\"asalmi to measure the geo-neutrino flux in a continental
location and another detector in Hawaii to measure it in an oceanic
location. The probability that only one of them is shadowed exceeds
50\% whereas the probability that at least one is shadowed is about 80\%.

As an important caveat, we mention that very recently it has been recognized that nonlinear oscillation effects caused by
neutrino-neutrino interactions can have a dramatic impact on the
neutrino flavor evolution for approximately the first 100~km above the
neutrino sphere \cite{Duan:2006an,Hannestad:2006nj}. 
The impact of these novel effects and of their observable signatures  is 
currently under investigation. However, from recent numerical simulations \cite{Duan:2006an}
and analytical studies \cite{Raffelt:2007cb}, it results that the effects of these non-linear 
effects would produce a spectral  swap $\nu_e \bar{\nu}_e \leftarrow \nu_x \bar{\nu}_x$
at $r \lesssim 400$~km, for inverted neutrino mass hierarchy. 
One would observe a complete spectral swapping, while $\nu$ spectra would show a 
peculiar bimodal split. These effect would appear also for
astonishingly small values of $\theta_{13}$. 
These new results suggests once more that one needs complementary detection
techniques to be sensitive to both neutrino and anti neutrino channels.

Other interesting ideas have been studied in the literature, as the pointing of a SN by neutrinos \cite{Tomas:2003xn},
determining its distance from the deleptonization burst that 
plays the role of a standard candle \cite{Kachelriess:2004ds},
an early alert for an SN observatory exploiting the neutrino
signal \cite{Antonioli:2004zb}, and the detection of neutrinos from
the last phases of a 
presupernova star \cite{Odrzywolek:2003vn}.

So far, we have investigated SN in our Galaxy, but the calculated
rate of supernova explosions within a distance of 10~Mpc is about 1/year. 
Although the number of events from a single explosion at
such large distances would be small, the signal could be separated from the background with the condition to observe at least 
two events within a time window comparable to the neutrino emission time-scale ($\sim 10$~sec), together with the full 
energy and time distribution of the events \cite{Ando:2005ka}. In the MEMPHYS detector, with at least
two neutrinos observed, a SN could be identified without optical confirmation, so that the start of the light curve could be
forecast by a few hours, along with a short list of probable host
galaxies. This would also allow the detection of supernovae which are either heavily obscured by dust or are optically 
faint due to prompt black hole formation.

\begin{table}
		\caption{\label{tab:Phys-SN-SummaryOscNeut}Summary
 of the effect of the neutrino properties on $\nu_e$ and $\bar{\nu}_e$ signals.}
		\begin{tabular}{@{}lllll}\br
		\parbox[b]{2cm}{Mass\\ Hierarchy}   & $\sin^2\theta_{13}$ & \parbox[b]{3cm}{$\nu_e$ neutronization\\peak} & Shock wave & Earth effect 
		\\
		\mr
		Normal    & $\gtrsim 10^{-3}$ & Absent  & $\nu_e$   & $\bar{\nu}_e$\\
		Inverted    & $\gtrsim 10^{-3}$ & Present  & $\bar{\nu}_e$   & $\nu_e$ \\
		Any    & $\lesssim 10^{-5}$ & Present  & -   & both $\bar{\nu}_e$ $\nu_e$ \\
\br
		\end{tabular}
\end{table}
\subsection{Diffuse supernova neutrino background} 

As mentioned above, a galactic SN explosion would be a spectacular source of neutrinos, 
so that a variety of neutrino and SN properties could be
assessed.  However, only one such explosion is expected in 20 to 100
years by now.  
Waiting for the next galactic SN, one can detect the cumulative neutrino flux from all the past SN in the Universe, 
the so-called Diffuse Supernova Neutrino Background (DSNB). In particular, there is an energy window around 
$10-40$~MeV where the DSNB signal can emerge above other sources, so that the proposed detectors may well 
measure this flux after some years of exposure.

\begin{table}
	\caption{\label{tab:Phys-SN-DiffuseRates}DSNB expected
	rates. The larger numbers of expected signal events are computed with the present limit
	on the flux by the Super-Kamiokande Collaboration. The smaller
	numbers are computed for typical models. The background
	from reactor plants has been computed for specific sites
	for LENA and MEMPHYS. For MEMPHYS, the Super-Kamiokande
	background has been scaled by the exposure.}

	\begin{tabular}{@{}llll}\br
	Interaction & Exposure     &  Energy Window &  Signal/Bkgd \\ \mr 
\multicolumn{4}{@{}l}{GLACIER}\\
 $\nu_e + {}^{40}Ar \rightarrow e^- + {}^{40}K^*$ &
\parbox{2cm}{0.5~Mton~year\\5~years} &
$[16-40]$~MeV & (40-60)/30 \\
\multicolumn{4}{@{}l}{LENA at Pyh\"asalmi} \\
\parbox{25mm}{$\bar{\nu}_e + p \rightarrow n + e^+$\\$n+p\rightarrow d+ \gamma$ (2~MeV, $200~\mu$s)} &
\parbox{2cm}{0.4~Mton~year\\10~years} & 
$[9.5-30]$~MeV & (20-230)/8 \\
\multicolumn{4}{@{}l}{1 MEMPHYS module + 0.2\% Gd (with bkgd at Kamioka)} \\
\parbox{3cm}{$\bar{\nu}_e + p \rightarrow n + e^+$\\$n+Gd\rightarrow \gamma$\\(8~MeV, $20~\mu$s)} &
\parbox{2cm}{0.7~Mton~year\\5~years} & 
$[15-30]$~MeV & (43-109)/47 \\
\br
		\end{tabular}
\end{table}
 
The DSNB signal, although weak, is not only  guaranteed, but can also allow 
probing physics different from that of a galactic SN, including
processes which occur on cosmological scales in time or space. 
For instance, the DSNB signal is sensitive to the evolution of the SN
rate, which in turn is closely related to the star formation rate
\cite{Fukugita:2002qw,Ando:2004sb}. In addition, neutrino decay
scenarios with cosmological lifetimes could be analyzed and
constrained \cite{Ando:2003ie} as proposed in \cite{Fogli:2004gy}.
An upper limit on the DSNB flux has been set by the Super-Kamiokande
experiment \cite{Malek:2002ns}

\begin{equation}
	\phi_{\bar{\nu}_e}^{\mathrm{DSNB}} < 1.2~ \flux (E_\nu > 19.3~\mathrm{MeV}).
\end{equation}

An upper limit based on the non observation of distortions of the expected
 background spectra in the same energy range. The most recent
 theoretical estimates  (see for example \cite{Strigari:2005hu,Hopkins:2006bw})  predict a DSNB flux very close to the SK upper limit,
 suggesting that the DSNB is on the verge of the detection if a
 significant background reduction is achieved such as Gd loading \cite{Beacom:2003nk}.  
 With a careful reduction of backgrounds, the proposed large detectors would
 not only be able to detect the DSNB, but to study its spectral
 properties with some precision.  In particular, MEMPHYS and LENA would be sensitive
 mostly to the $\bar{\nu}_e$ component of DSNB,  through $\bar{\nu}_e$ IBD, 
 while GLACIER would probe  $\nu_e$ flux, trough   $\nu_e + {}^{40}Ar     
\rightarrow e^-  + {}^{40}K^*$ (and the  associated gamma cascade) \cite{Cocco:2004ac}.

\begin{figure}
\begin{center}
\includegraphics[width=0.7\columnwidth]{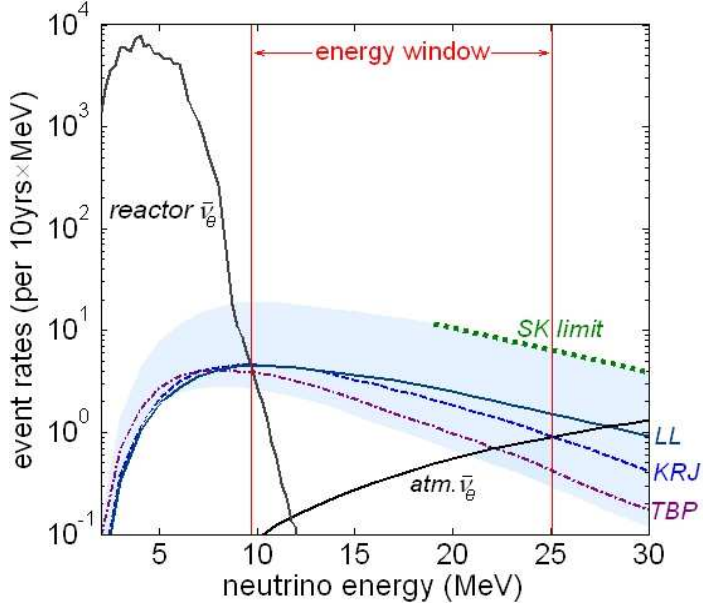}
\end{center}
\caption{DSNB signal and background in the LENA detector in 10 years of exposure. The shaded regions give the uncertainties of all curves. An observational window between $\sim 9.5$ to 25~MeV that is almost free of background can be identified 
(for the Pyh\"asalmi site). Reprinted figure with permission from~\cite{Wurm:2007cy}.}
\label{fig:Phys-SN-LENAsnr}
\end{figure}

\begin{figure}
\begin{center}
\includegraphics[width=0.7\columnwidth]{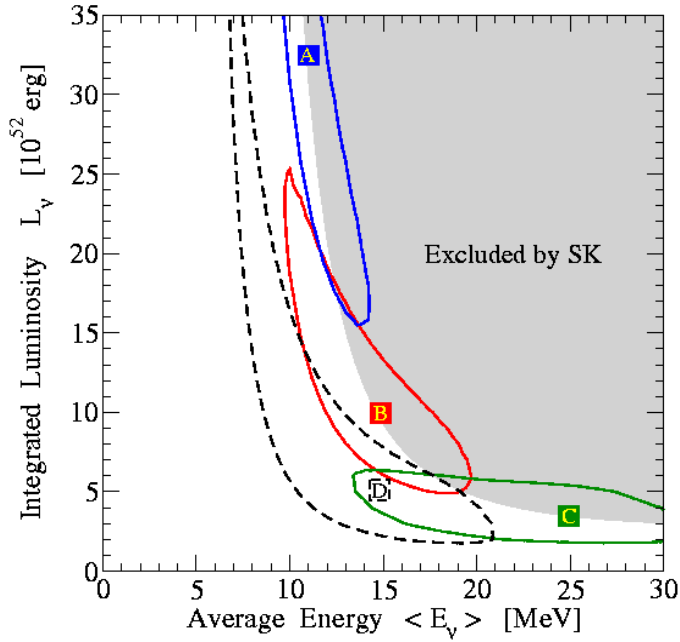}
\end{center}
\caption{Possible 90\% C.L. measurements of the emission parameters
of supernova electron antineutrino emission after 5
years running of a Gadolinium-enhanced SK detector or 1 year of one Gadolinium-enhanced MEMPHYS tanks.
Reprinted figure with permission from~\cite{Yuksel:2005ae}.}
\label{fig:Phys-DSN-sndpar}
\end{figure}

The DSNB signal energy window is constrained from above by the atmospheric neutrinos and from below by 
either the nuclear reactor $\bar{\nu}_e$ (I), the spallation production of unstable radionuclei
by cosmic-ray muons (II), the decay of "invisible" muons into electrons (III), solar
 $\nu_e$ neutrinos (IV), and low energy atmospheric $\nu_e$ and $\bar{\nu}_e$ neutrinos interactions (V). The three detectors
are affected differently by these backgrounds.
GLACIER looking at $\nu_e$ is mainly affected by types IV and V. MEMPHYS filled with pure water is affected by types I, II, V and III due to the 
fact that the muons may not have enough energy to produce Cherenkov light. As pointed out in \cite{Fogli:2004ff}, with the addition of Gadolinium \cite{Beacom:2003nk} the detection of the captured neutron releasing 8~MeV gamma after 
$\sim20~\mu$s (10 times faster than in pure water) would give the possibility to reject  the "invisible" muon (type III) 
as well as the spallation background (type II). 
LENA taking benefit from the delayed neutron capture in $\bar{\nu}_e + p \rightarrow n + e^+$, is mainly concerned with
reactor neutrinos (I), which impose to choose an underground site far from nuclear plants.
If LENA was installed at the Center for Underground Physics in Pyh\"asalmi (CUPP, Finland),
there would be an observational window from $\sim 9.7$ to 25~MeV that is almost free of background. The expected rates of signal and background are presented in \refTab{tab:Phys-SN-DiffuseRates}.
According to current DSNB models \cite{Ando:2004sb} that are using
different SN simulations (\cite{Totani:1997vj, Thompson:2002mw, Keil:2002in}) for the
prediction of the DSNB energy spectrum and flux, the detection of $\sim$10 DSNB events per year is realistic for LENA. Signal rates
corresponding to different DSNB models and the background rates due to reactor and atmospheric neutrinos are shown in
\refFig{fig:Phys-SN-LENAsnr} for 10 years exposure at CUPP.

Apart from the mere detection, spectroscopy of DSNB events in LENA will constrain the parameter space of core-collapse models. 
If the SN rate signal is known with sufficient precision, the spectral slope of the DSNB can be used to determine 
the hardness of the initial SN neutrino spectrum. For the currently favoured value of the SN rate, the discrimination between core-collapse models will be possible at 2.6$\sigma$ after 10 years of measuring time \cite{Wurm:2007cy}.
In addition, by the analysis of the flux in the energy region from 10
to 14~MeV the SN rate for $z<2$ could be constrained with high significance, as in this energy regime the DSNB flux is only weakly dependent on the assumed SN model. 
The detection of the redshifted DSNB from $z>1$ is limited by the flux of the reactor $\bar\nu_e$ background. In Pyhäsalmi, a lower threshold of 9.5~MeV resuls in a spectral contribution of 25\% DSNB from $z>1$.

The analysis of the expected DSNB spectrum that would be observed
with a Gadolinium-loaded \WC\ detector has been carried out in \cite{Yuksel:2005ae}. 
The possible measurements of the parameters (integrated luminosity and average energy) of
SN $\bar\nu_e$ emission have been computed for 5 years running of
a Gd-enhanced Super-Kamiokande detector, which would correspond to 1 year
of one Gd-enhanced MEMPHYS tank. The results are shown in \refFig{fig:Phys-DSN-sndpar}.
Even if detailed studies on the characterization of the background are needed, the DSNB events provide the first neutrino detection originating from cosmological distances.

\section{Solar neutrinos}
\label{sec:Solar}
In the past years water Cherenkov detectors have measured the high energy tail ($E>5$~MeV)
of the solar $^{8}$B neutrino flux using electron-neutrino elastic scattering \cite{Smy:2002rz}. 
Since such detectors could record the time of an interaction and reconstruct 
the energy and direction of the recoiling electron, unique information 
on the spectrum and time variation of the solar neutrino flux were extracted. 
This provided further insights into the "solar neutrino problem'', 
the deficit of the neutrino flux (measured by several experiments) 
with respect to the flux expected by solar models, contributing to the assessment of
the oscillation scenario for solar neutrinos \cite{Davis:1968cp,Hirata:1989zj,Anselmann:1992um,Abdurashitov:1994bc,Smy:2002rz,Aharmim:2005gt,Altmann:2005ix} .

With MEMPHYS,  Super-Kamiokande's measurements obtained from 1258 days 
of data taking could be repeated in about half a year, while the seasonal flux variation 
measurement will obviously require a full year. In particular, the first 
measurement of the flux of the rare $hep$ neutrinos may be possible.
Elastic neutrino-electron scattering is strongly forward peaked. 
In order to separate the solar neutrino signal from the isotropic background events (mainly due to low radioactivity), this 
directional correlation is exploited, although the angular resolution is limited 
by multiple scattering.  The reconstruction algorithms first reconstruct 
the vertex from the PMT timing information and then the direction, by assuming a single 
Cherenkov cone originating from the reconstructed vertex. 
Reconstructing 7~MeV events in MEMPHYS seems not to be a problem, but decreasing this 
threshold would imply serious consideration of the PMT dark current rate as well as the laboratory and detector radioactivity level.

With LENA, a large amount of neutrinos from ${}^{7}$Be (around $\sim5.4\times10^3$/day, $\sim 2.0\times10^6$/year) would be
detected. Depending on the signal to background ratio, this could provide a sensitivity to time variations in the $^{7}$Be neutrino
flux of $\sim 0.5$\% during one month of measuring time. Such a sensitivity can give unique information on helioseismology
(pressure or temperature fluctuations in the center of the Sun) and on a possible magnetic moment interaction 
with a timely varying solar magnetic field.
The {\it pep} neutrinos are expected to be recorded at a
rate of $210$/day ($\sim 7.7\times10^4$/y). These events would
provide a better understanding of the global solar neutrino
luminosity, allowing to probe (due to their peculiar energy)  the
transition region of vacuum to matter-dominated neutrino oscillation.

The neutrino flux from the CNO cycle is theoretically
predicted with a large uncertainty (30\%). Therefore, LENA would provide a new opportunity for a detailed
study of solar physics. However, the observation of such solar
neutrinos in these detectors, $i.e.$ through elastic scattering, is not
a simple task, since neutrino events cannot be separated from the background, and it can be accomplished only if the detector
contamination will be kept very low \cite{Alimonti:1998aa,Alimonti:1998nt}. Moreover, only
mono-energetic sources as those mentioned can be detected, taking
advantage of the Compton-like shoulder edge produced in the event spectrum.

Recently, the possibility to detect ${}^8$B solar neutrinos by means of charged-current interaction with the
${}^{13}$C \cite{Ianni:2005ki} nuclei naturally contained in organic scintillators has been investigated. Even if signal events do not
keep the directionality of the neutrino, they can be separated from background by exploiting the time and space coincidence with the
subsequent decay of the produced ${}^{13}$N nuclei. The residual background amounts to about $~60$/year 
corresponding to a reduction factor of
$\sim 3 \times10^{-4}$) \cite{Ianni:2005ki}. Around 360~events of this type
per year can be estimated for LENA. A deformation due to the MSW matter effect
should be observable in the low-energy regime after a couple of years of measurements.

For the proposed location of LENA in Pyh\"asalmi ($\sim 4000$~m.w.e.),
the cosmogenic background will be sufficiently low for the above mentioned
measurements. Notice that the Fréjus site would also be adequate for this
case ($\sim 4800$~m.w.e.). The radioactivity of the detector would
have to be kept very low ($10^{-17}$~g/g level U-Th) as in the KamLAND detector.

Solar neutrinos can be detected by GLACIER through the elastic scattering $\nu_x + e^- \rightarrow \nu_x + e^-$ (ES) and the absorption 
reaction $\nu_e + {}^{40}Ar \rightarrow e^- + {}^{40}K^*$ (ABS) followed by $\gamma$-ray emission. 
Even if these reactions have low energy threshold ($1.5$~MeV for the second one), 
one expects to operate in practice with a threshold set at 5~MeV on the primary electron kinetic energy,
in order to reject background from neutron capture followed by gamma emission, which constitutes the main background for some
of the underground laboratories \cite{Arneodo:2001tx}. 
These neutrons are induced by the spontaneous fission and ($\alpha$,n)
reactions in rock. In the case of a salt mine this background can be smaller.
The fact that salt has smaller U/Th concentrations does not necessarily mean that the neutron flux is smaller. The flux depends on the rock
composition since (alpha,n) reactions may contribute significantly to the flux.
The expected raw event rate is $330\ 000$/year (66\% from ABS, 25\% from ES and 9\% from neutron background induced events) 
assuming the above mentioned threshold on the final electron energy. 
By applying further offline cuts to purify separately the ES sample and the ABS sample, one obtains 
the rates shown on \refTab{tab:GLACIER-Solar}.

\begin{table}
		\caption{\label{tab:GLACIER-Solar} Number of events expected in GLACIER per year, compared with the computed background (no oscillation) from the Gran Sasso rock radioactivity ($0.32~10^{-6}$~n \flux ($> 2.5$~MeV). The absorption channel has 
been split into the contributions of events from Fermi and Gamow-Teller transitions of the ${}^{40}$Ar to the different ${}^{40}$K excited levels and that can be separated using the emitted gamma energy and multiplicity.} 
\lineup
\begin{indented}
\item[]\begin{tabular}{@{}ll}\br
							& Events/year \\ \mr
Elastic channel ($E\geq5$~MeV)                &   $\045\ 300$ \\
Neutron background          						      &	  $\0\0\ 1400$ \\
Absorption events contamination               & $\0\0\ 1100$ \\ \mr
Absorption channel (Gamow-Teller transition)	& $101\ 700$ \\
Absorption channel (Fermi transition)	        & $\059\ 900$ \\
Neutron background          						      & $\0\0\ 5500$ \\						
Elastic events contamination                  & $\0\0\ 1700$ \\		
			\br
		\end{tabular}
		\end{indented}
\end{table}

A possible way to combine the ES and the ABS channels similar to the NC/CC flux ratio measured by SNO collaboration \cite{Aharmim:2005gt}, is to compute the following ratio

\begin{equation}
	R = \frac{N^{ES}/N^{ES}_0}{\frac{1}{2}\left( N^{Abs-GT}/N^{Abs-GT}_0 + N^{Abs-F}/N^{Abs-F}_0\right)}
\end{equation}

where the numbers of expected events without neutrino oscillations are labeled with a $0$). 
This double ratio has two advantages.
First, it is independent of the ${}^{8}$B total neutrino flux, predicted by different solar models, 
and second, it is free from experimental threshold energy bias and of the adopted cross-sections 
for the different channels. 
With the present fit to solar neutrino experiments and KamLAND data, one expects a value of $R = 1.30\pm 0.01$ after one 
year of data taking with GLACIER.  The quoted error for R only takes into account statistics. 

\section{Atmospheric neutrinos}
\label{sec:Phys-Atm-neut}

Atmospheric neutrinos originate from the decay chain initiated by the collision of 
primary cosmic-rays with the upper layers of Earth's atmosphere.
The primary cosmic-rays are mainly protons
and helium nuclei producing secondary particles such
$\pi$ and $K$, which in turn decay producing electron- and muon-
neutrinos and antineutrinos.

\begin{figure}
\begin{center}
    \includegraphics[width=0.7\columnwidth]{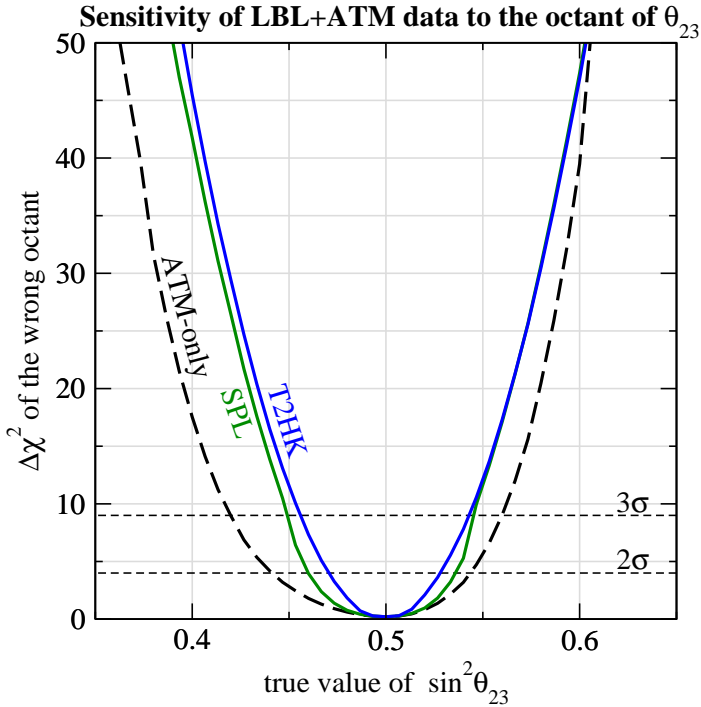}
\end{center}
    \caption{ \label{fig:octant} %
      Discrimination of the wrong octant solution as a function of
      $\sin^2\theta_{23}^\mathrm{true}$, for
      $\theta_{13}^\mathrm{true} = 0$. We have assumed 10 years of
      data taking with a 440 kton detector. Reprinted figure with permission from~\cite{Campagne:2006yx}.}
\end{figure}

At low energies the main contribution comes from $\pi$ mesons, and
the decay chain $\pi \to \mu + \nu_\mu$ followed by $\mu \to e + \nu_e
+ \nu_\mu$ produces essentially two $\nu_\mu$ for each $\nu_e$.  As
the energy increases, more and more muons reach the ground before
decaying, and therefore the $\nu_\mu / \nu_e$ ratio increases.
For $E_\nu \gtrsim 1$~GeV the dependence of the total neutrino flux on
the neutrino energy is well described by a power law, $d\Phi / dE
\propto E^{-\gamma}$ with $\gamma = 3$ for $\nu_\mu$ and $\gamma=3.5$
for $\nu_e$, whereas for sub-GeV energies the dependence becomes more
complicated because of the effects of the solar wind and of Earth's magnetic field \cite{Gonzalez-Garcia:2002dz}. As for the
zenith dependence, for energies larger than a few GeV the neutrino
flux is enhanced in the horizontal direction, since pions and muons can travel a longer distance before
losing energy in interactions (pions) or reaching the ground (muons),
and therefore have more chances to decay producing energetic neutrinos.

Historically, the atmospheric neutrino problem originated in the 80's as a discrepancy between the 
atmospheric neutrino flux measured
with different experimental techniques and the expectations. In the last years, a
number of detectors had been built, which could detect neutrinos through the observation of the charged lepton produced in charged-current neutrino-nucleon interactions inside the detector material.
These detectors could be divided into two classes: \emph{iron calorimeters}, which reconstruct the track or the 
electromagnetic shower induced by the lepton, and \emph{water Cherenkov}, which measure the Cherenkov light 
emitted by the lepton as it moved faster
than light in water filling the detector volume.
The first iron calorimeters, Frejus \cite{Daum:1994bf} and NUSEX \cite{Aglietta:1988be}, found no discrepancy between the
observed flux and the theoretical predictions, whereas the two \WC\ detectors, IMB \cite{Becker-Szendy:1992hq} and
Kamiokande \cite{Hirata:1992ku}, observed a clear deficit compared to the predicted $\nu_\mu / \nu_e$ ratio.
The problem was finally solved in 1998, when the already mentioned water Cherenkov
Super-Kamiokande detector \cite{Fukuda:1998mi} allowed to establish with high
statistical accuracy that there was indeed a zenith- and energy-dependent deficit in the muon-neutrino flux with respect to the
theoretical predictions, and that this deficit was compatible with the
hypothesis of  $\nu_\mu \to \nu_\tau$ oscillations. The independent confirmation of this effect from the calorimeter
experiments Soudan-II \cite{Allison:1999ms} and
MACRO \cite{Ambrosio:2001je} eliminated the original discrepancy between the
two experimental techniques.

Despite providing the first solid evidence for neutrino oscillations,
atmospheric neutrino experiments suffer from two important limitations.
Firstly, the sensitivity of an atmospheric neutrino experiments is
strongly limited by the large uncertainties in the knowledge of
neutrino fluxes and neutrino-nucleon cross-section. Such uncertainties can be as large as 20\%.
Secondly, water Cherenkov detectors do not allow an accurate
    reconstruction of the neutrino energy and direction if none of the
    two is known a priori. This strongly limits the sensitivity to
    $\Delta m^2$, which is very sensitive to the resolution of $L/E$.

During its phase-I, Super-Kamiokande has collected 4099 electron-like
and 5436 muon-like contained neutrino events \cite{Ashie:2005ik}. With
only about one hundred events each, the accelerator experiments K2K \cite{Ahn:2006zz} and
MINOS \cite{Tagg:2006sx} already provide a stronger bound on the atmospheric mass-squared difference $\Delta m_{31}^2$. The present
value of the mixing angle $\theta_{23}$ is still dominated by Super-Kamiokande data, being statistically the most important factor for
such a measurement. However, large improvements are expected from the next
generation of long-baseline experiments such as T2K \cite{Itow:2001ee} and
NO$\nu$A \cite{Ayres:2004js}, sensitive to the same oscillation parameters as atmospheric neutrino experiments.

\begin{figure}
\begin{center}
    \includegraphics[width=0.7\columnwidth]{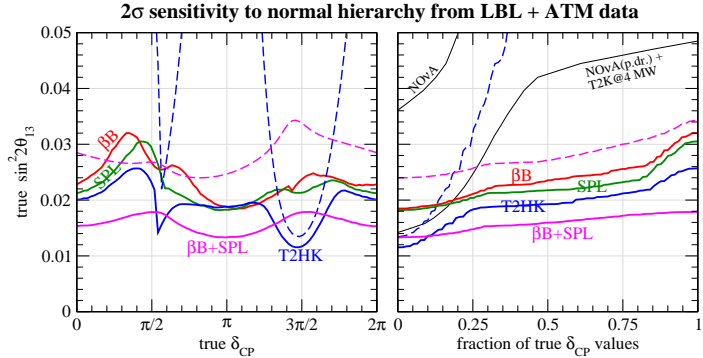}
\end{center}
    \caption{ \label{fig:hierarchy} %
      Sensitivity to the mass hierarchy at $2\sigma$ ($\Delta\chi^2 =
      4$) as a function of $\sin^22\theta_{13}^\mathrm{true}$ and
      $\delta_\mathrm{CP}^\mathrm{true}$ (left), and the fraction of
      true values of $\delta_\mathrm{CP}^\mathrm{true}$ (right). The
      solid curves are the sensitivities from the combination of
      long-baseline and atmospheric neutrino data, the dashed curves
      correspond to long-baseline data only. We have assumed 10 years
      of data taking with a 440~kton mass detector. Reprinted figure with permission from~\cite{Campagne:2006yx}.}
\end{figure}

\begin{figure}
\begin{center}
    \includegraphics[width=0.7\columnwidth]{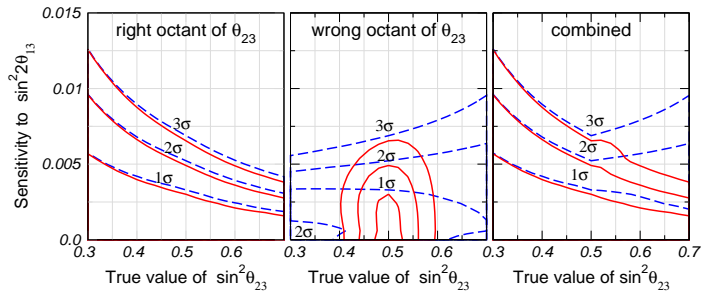}
\end{center}
    \caption{ \label{fig:theta13} %
      Sensitivity to $\sin^22\theta_{13}$ as a function of
      $\sin^2\theta_{23}^\mathrm{true}$ for LBL data only (dashed),
      and the combination beam and atmospheric neutrino data (solid). In the left and central
      panels we restrict the fit of $\theta_{23}$ to the octant
      corresponding to $\theta_{23}^\mathrm{true}$ and $\pi/2 -
      \theta_{23}^\mathrm{true}$, respectively, whereas the right
      panel shows the overall sensitivity taking into account both
      octants. We have assumed 8 years of beam and 9 years of atmospheric neutrino data
      taking with the T2HK beam and a 1~Mton detector. Reprinted figure with permission from~\cite{huber-2005-71}.}
\end{figure}

Despite the above limitations, atmospheric neutrino detectors can still play a leading role in the future of neutrino physics due to the huge range
in energy (from 100~MeV to 10~TeV and above) and distance (from 20~km to more than $12\ 000$~km) covered by the data. 
This unique feature, as well as the very large statistics expected for a detector such as
MEMPHYS ($20\div 30$ times the present Super-Kamiokande event rate), will allow a
very accurate study of the subdominant modification to the leading
oscillation pattern, thus providing complementary information to
accelerator-based experiments. More concretely, atmospheric neutrino
data will be extremely valuable for
\begin{itemize}
  \item Resolving the octant ambiguity. Although future accelerator
    experiments are expected to considerably improve the measurement
    of the absolute value of the small quantity $D_{23} \equiv
    \sin^2\theta_{23} - 1/2$, they will have practically no
    sensitivity on its sign.  On the other hands, it has been pointed
    out \cite{Kim:1998bv,Peres:1999yi} that the $\nu_\mu \to \nu_e$ conversion
    signal induced by the small but finite value of $\Delta m_{21}^2$
    can resolve this degeneracy. However, observing such a conversion
    requires a very long baseline and low energy neutrinos, and
    atmospheric sub-GeV electron-like events are particularly suitable
    for this purpose. In \refFig{fig:octant} we show the potential
    of different experiments to exclude the octant degenerate
    solution.

  \item Resolving the hierarchy degeneracy. If $\theta_{13}$ is not
    too small, matter effect will produce resonant conversion in the
    $\nu_\mu \leftrightarrow \nu_e$ channel for neutrinos
    (antineutrinos) if the mass hierarchy is normal (inverted). The
    observation of this enhanced conversion would allow the
    determination of the mass hierarchy. Although a magnetized
    detector would be the best solution for this task, it is possible
    to extract useful information also with a conventional detector
    since the event rates expected for atmospheric neutrinos and
    antineutrinos are quite different. This is clearly visible from
    \refFig{fig:hierarchy}, where we show how the sensitivity to the
    mass hierarchy of different beam experiments is drastically
    increased when the atmospheric neutrino data collected by the same detector are
    also included in the fit.

  \item Measuring or improving the bound on $\theta_{13}$. Although
    atmospheric data alone are not expected to be competitive with the
    next generation of long-baseline experiments in the sensitivity to
    $\theta_{13}$, they will contribute indirectly by eliminating the
    octant degeneracy, which is an important source of uncertainty for beam experiments.
    In particular, if $\theta_{23}^\mathrm{true}$ is larger than
    $45^\circ$ then the inclusion of atmospheric data will
    considerably improve the accelerator experiment sensitivity to $\theta_{13}$, as can
    be seen from the right panel of \refFig{fig:theta13} \cite{huber-2005-71}.
\end{itemize}


In GLACIER, the search for $\nu_\tau$ appearance is based on the information provided by the event kinematics and takes advantage of the special characteristics of $\nu_\tau$ CC and the subsequent
decay of the produced $\tau$ lepton when compared to CC and NC interactions 
of $\nu_\mu$ and $\nu_e$, i.e. by making use of $\vec{P}_{candidate}$ 
and $\vec{P}_{hadron}$. 
Due to the large background induced by atmospheric muon and electron
neutrinos and antineutrinos, the measurement of a statistically
significant excess of $\nu_\tau$ 
events is very unlikely for the  $\tau \to e$ and  $\tau \to \mu$ decay modes.

The situation is much more advantageous for the hadronic channels. 
One can consider tau-decays to one prong (single pion, $\rho$) and to three
prongs ($\pi^\pm \pi^0 \pi^0 $ and three charged pions). In order to select the signal, 
one can exploit the kinematical variables $E_{visible}$, 
$y_{bj}$ (the ratio between the total hadronic energy and
$E_{visible}$) and $Q_T$ (defined as the transverse momentum of the $\tau$
candidate with respect to the total measured momentum) that are not completely independent one from another but show
some correlation. These correlations can be exploited to reduce the
background. In order to maximize the separation between signal 
and background, one can use three dimensional likelihood functions 
${\cal L}(Q_T,E_{visible}, y_{bj})$ where
correlations are taken into account. For each channel, three
dimensional likelihood functions are built 
for both signal (${\cal L}^S_\pi, \ {\cal L}^S_\rho, \ 
{\cal L}^S_{3\pi}$) and background (${\cal L}^B_\pi, \ {\cal L}^B_\rho, \ 
{\cal L}^B_{3\pi}$). In order to enhance the separation of $\nu_\tau$ induced 
events from $\nu_\mu, \ \nu_e$ interactions, the ratio of 
likelihoods is taken as the sole discriminant variable
$\ln \lambda_i \equiv \ln({\cal L}^S_i / {\cal L}^B_i)$ where $i=\pi,\ \rho, \ 3\pi$.

To further improve the sensitivity of the $\nu_\tau$ appearance search, one can combine
the three independent hadronic analyses into a single one. Events that are common to at least 
two analyses are counted only once and a survey of all possible combinations, for a restricted set of  values of the likelihood 
ratios, is performed. Table \ref{tab:combi} illustrates the  statistical significance achieved by several selected combinations of the 
likelihood ratios for an exposure equivalent to 100 kton year. 

\begin{table}
\caption{\label{tab:combi}Expected GLACIER background and signal events for different
combinations of the $\pi$, $\rho$ and $3\pi$ analyses. The considered
statistical sample corresponds to an exposure of 100
kton year.}
\lineup
\begin{indented}
\item[]\begin{tabular}{@{}llllll}\br
$\ln \lambda_\pi$ & $\ln \lambda_\rho$ & $\ln \lambda_{3\pi}$ & 
Top & Bottom & $P_\beta$ ($\%$) \\
Cut & Cut & Cut & Events & Events &  \\ \mr
0.0 & $\m0.5$ & $\m 0.0$ & $223$ & $223 + 43 = 266$ & $2 \times 10^{-1}$ ($3.1\sigma$)\\
1.5 & $\m1.5$ & $\m 0.0$ & $\0 92$ & $\0 92 + 35= 127$ & $2 \times 10^{-2}$ ($3.7\sigma$)\\
3.0 & $-1.0$ & $\m 0.0$ & $\0 87$ & $\0 87 + 33 = 120 $ & $3 \times 10^{-2}$ ($3.6\sigma$)\\
3.0 & $\m0.5$ & $\m 0.0$ & $\0 25$ & {$\0 25 + 22= 47$} & {$2 \times 10^{-3}$ $(4.3\sigma)$} \\ 
3.0 & $\m1.5$ & $\m 0.0$ & $\0 20$ & $\0 20 + 19 = 39$ & $4 \times 10^{-3}$ ($4.1\sigma$)\\
3.0 & $\m0.5$ & $-1.0$ & $\0 59$ & $\0 59 + 30 = 89$ & $9 \times 10^{-3}$ ($3.9\sigma$)\\
3.0 & $\m0.5$ & $\m 1.0$ & $\0 18$ & $\0 18 + 17 = 35$ & $1 \times 10^{-2}$ ($3.8\sigma$)\\ \br
\end{tabular}
\end{indented}
\end{table}

The best combination for a 100 kton year exposure is achieved for the 
following set of cuts: {$\ln \lambda_\pi > 3$, $\ln \lambda_\rho > 0.5$} and {$\ln \lambda_{3\pi} > 0$}. 
The expected number of NC background events amounts to 25 (top) 
while 25+22 = 47 are expected. $P_\beta$ is the Poisson probability 
for the measured excess of upward going events to be due to a
statistical fluctuation as a function of the exposure. An effect larger than $4\sigma$ is obtained for an 
exposure of 100 kton year (one year of data taking with GLACIER).

Last but not least, it is worth noting that atmospheric neutrino fluxes are
themselves an important subject of investigation, and in the light of
the precise determination of the oscillation parameters provided by
long baseline experiments, the atmospheric neutrino data accumulated by
the proposed detectors could be used as a direct measurement of the incoming
neutrino flux, and therefore as an indirect measurement of the primary cosmic-rays flux. 

The appearance  of subleading features in the main oscillation pattern can also be
    a hint for New Physics. The huge range of energies probed by
    atmospheric data will allow to set very strong bounds on
    mechanisms which predict deviation from the $1/E$ law behavior. For
    example, the bound on non-standard neutrino-matter interactions
    and on other types of New Physics (such as violation of the
    equivalence principle, or violation of the Lorentz invariance)
    which can be derived from present data is already the
    strongest which can be put on these
    mechanisms \cite{Gonzalez-Garcia:2004wg}. 

\section{Geo-neutrinos}
\label{sec:Geo}

The total power dissipated from the Earth (heat flow) has been
measured with thermal techniques to be $44.2\pm1.0$~TW. Despite this
small quoted error, a more recent evaluation of the same data
(assuming much lower hydrothermal heat flow near mid-ocean ridges) has
led to a lower figure of $31\pm1$~TW.
On the basis of studies of
chondritic meteorites the calculated radiogenic power is thought to be
19~TW (about half of the total power), 84\% of which is produced by
${}^{238}$U and ${}^{232}$Th decay which in turn produce $\bar{\nu}_e$
by beta-decays (geo-neutrinos).
It is then of prime importance to measure the
$\bar{\nu}_e$ flux coming from the Earth to get geophysical
information, with possible applications in the interpretation of the geomagnetism.

The KamLAND collaboration has recently reported the first observation
of the geo-neutrinos \cite{Araki:2005qa}. The events are identified by
the time and distance coincidence between the prompt $e^+$ and the
delayed (200~$\mu$s) neutron capture produced by $\bar{\nu}_e + p
\rightarrow n + e^+$ and emiting a 2.2~MeV gamma. The energy window
to search for the geo-neutrino events is $[1.7,3.4]$~MeV. The lower bound
corresponds to the reaction threshold while the upper bound is
constrained by nuclear reactor induced background events.
The measured rate in the 1~kton liquid scintillator detector located at
the Kamioka mine, where the Kamiokande detector was previously installed,
is $25^{+19}_{-18}$ for a total background of $127\pm 13$ events. 

The background is composed by $2/3$ of $\bar{\nu}_e$ events from
the nuclear reactors in Japan and Korea.
These events have been actually used by KamLAND to confirm and precisely measure the Solar driven
neutrino oscillation parameters (see Section \ref{sec:Solar}).
The residual $1/3$ of the events originates
from neutrons of 7.3~MeV produced in ${}^{13}$C$(\alpha,n){}^{16}$O reactions and captured as in the
IBD reaction. 
The $\alpha$ particles come from the ${}^{210}$Po decays, a ${}^{222}$Rn daughter which is of natural
radioactivity origin.  The measured geo-neutrino events can be
converted in a rate of $5.1^{+3.9}_{-3.6} \times 10^{-31}$ $\bar{\nu}_e$ per
target proton per year corresponding to a mean flux of
$5.7 \times 10^{6}\flux$, or this can be transformed into a $99\%$ C.L. upper
bound of $1.45 \times 10^{-30}$ $\bar{\nu}_e$ per target proton per year
($1.62 \times 10^{7}\flux$ and 60~TW for the radiogenic power).

In MEMPHYS, one expects 10 times more geo-neutrino events but this would imply to decrease the trigger 
threshold to 2~MeV which seems very challenging with respect to the present Super-Kamiokande threshold, set to 
4.6~MeV due to high level of raw trigger rate \cite{Fukuda:2002uc}. 
This trigger rate is driven by a number of factors as dark current of the 
PMTs, $\gamma$s from the rock surrounding the detector, radioactive decay in the PMT glass itself and Radon 
contamination in the water. 

In LENA at CUPP a geo-neutrino rate of
roughly 1000/year~\cite{Hochmuth:2005nh} from the dominant $ \bar\nu_e+p\to
e^+ + n $ IBD reaction is expected. The delayed
coincidence measurement of the positron and the 2.2 MeV gamma event, following neutron capture on protons in 
the scintillator provides a very efficient tool to reject background events.
The threshold energy of 1.8 MeV allows the measurement of geo-neutrinos
from the Uranium and Thorium series, but not from $^{40}$K.
A reactor background rate of about 240 events per year for LENA at CUPP in the relevant energy window from 1.8~MeV to
3.2~MeV has been calculated.
This background can be subtracted statistically using the information
on the entire reactor neutrino spectrum up to $\simeq$~8 MeV.  

As it was shown in KamLAND, a serious background source may come from radio
impurities. There the correlated background from the isotope
$^{210}$Po is dominating. However, with an enhanced radiopurity of the
scintillator, the background can be significantly reduced. 
Taking the radio purity levels of the Borexino CTF detector
at Gran Sasso, where a $^{210}$Po activity
of $35\pm12/\rm{m^3 day}$ in PXE has been observed, this background would
be reduced by a factor of about 150 compared to KamLAND and would
account to less than 10 events per year in the LENA detector.  

An additional background that fakes the geo-neutrino signal is due to
$^9$Li, which is produced by cosmic-muons in spallation reactions with
$^{12}$C and decays in a $\beta$-neutron cascade.  
Only a small part of the $^9$Li decays falls into the energy window which is relevant
for geo-neutrinos. KamLAND estimates this background to be $0.30 \pm
0.05$ \cite{Araki:2005qa}.

At CUPP the muon reaction rate would be
reduced by a factor $\simeq 10$ due to better shielding and this
background rate should be at the negligible level of $\simeq$~1 event per year in LENA.
From these considerations it follows that LENA would be a very capable
detector for measuring geo-neutrinos.  Different Earth models could
be tested with great significance. The sensitivity of LENA for probing
the unorthodox idea of a geo-reactor in the Earth's core was estimated,
too. At the CUPP underground laboratory the neutrino
background with energies up to $\simeq 8$~MeV due to nuclear power
plants was calculated to be around 2200 events per year.  A
2~TW geo-reactor in the Earth's core would contribute 420 events per
year and could be identified at a statistical level of better than
$3\sigma$ after only one year of measurement.

Finally, in GLACIER the $\bar{\nu}_e + {}^{40}Ar \rightarrow e^+ + {}^{40}Cl^*$ has a threshold 
of $~7.5$~MeV, which is too high for geo-neutrino detection.

\section{Indirect searches for the Dark Matter of the Universe}
\label{sec:DM}

The Weakly Interacting Massive Particles (WIMPs) that likely
constitute the halo of the Milky Way can occasionally interact with massive objects, 
such as stars or planets. When they scatter off such an object, 
they can potentially lose enough energy that they become gravitationally bound and 
eventually will settle in the center of the celestial body. In
particular, WIMPs can be captured by and accumulate in the core of the Sun. 

\begin{figure}
\begin{center}
\includegraphics[width=0.7\columnwidth]{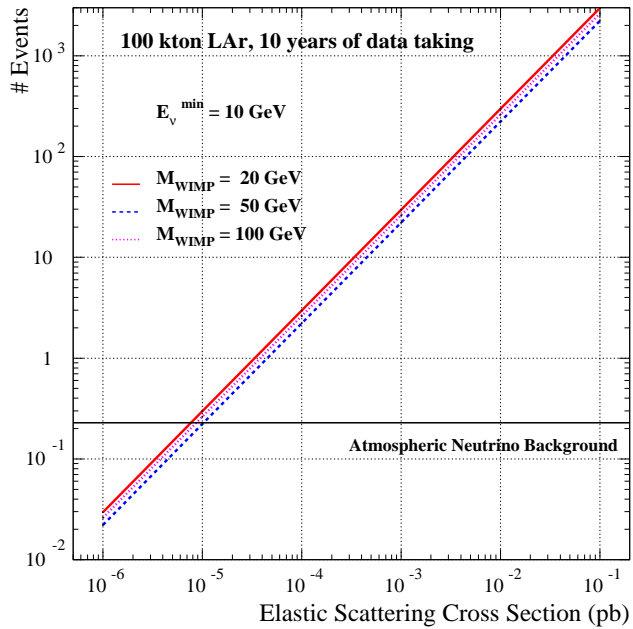}
\end{center}
\caption{\label{fig:GLACIERdm1} 
Expected number of signal and background events as a function of the
 WIMP elastic scattering production cross-section in the Sun, with a cut 
of 10 GeV on the minimum neutrino energy. Reprinted figure with permission from~\cite{Bueno:2004dv}.} 
\end{figure}

\begin{figure}
\begin{center}
\includegraphics[width=0.7\columnwidth]{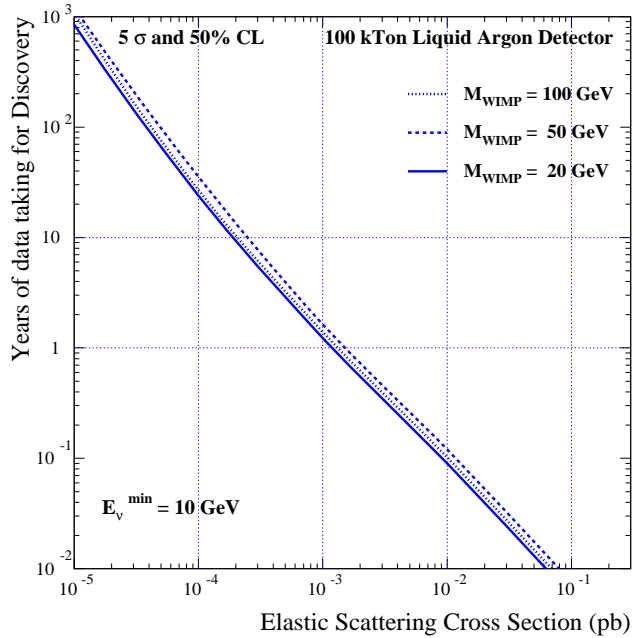}
\end{center}
\caption{\label{fig:GLACIERdm2} Minimum number of years required to claim a discovery WIMP signal
 from the Sun in a 100~kton LAr detector as function of $\sigma_{\rm{elastic}}$
 for three values of the WIMP mass. Reprinted figure with permission from~\cite{Bueno:2004dv}.}
\end{figure}

As far as the next generation of large underground observatories is concerned, although not specifically dedicated to the
search for WIMP particles, one could discuss the capability of GLACIER in identifying, 
in a model-independent way, 
neutrino signatures coming from the products of WIMP annihilations in the core 
of the Sun \cite{Bueno:2004dv}. 

Signal events will consist of energetic electron- (anti)neutrinos coming from the decay
of $\tau$ leptons and $b$ quarks produced in WIMP annihilation in 
the core of the Sun. Background contamination from atmospheric neutrinos is expected to be low. 
One cannot consider the possibility of observing neutrinos from WIMPs accumulated in the Earth. 
Given the smaller mass of the Earth and the fact that only scalar interactions contribute, 
the capture rates for our planet are not enough to produce a statistically
significant signal in GLACIER.

The search method takes advantage of the excellent angular reconstruction and 
superb electron identification capabilities GLACIER offers in looking for an excess of 
energetic electron- (anti)neutrinos pointing in the direction of the
Sun. The expected signal and background event rates have been evaluated, as said above in
a model independent way, as a function of the WIMP elastic scattering cross-section for a range of masses up to 100~GeV.
The detector discovery potential, namely the number of years needed to
claim a WIMP signal has been discovered, is shown in Figs.~\ref{fig:GLACIERdm1} 
and \ref{fig:GLACIERdm2}. With the assumed set-up and thanks to the low background environment 
provided by the LAr TPC, a clear WIMP signal would be detected
provided the elastic scattering cross-section in the Sun is above $\sim 10^{-4}$~pb.

\section{Neutrinos from nuclear reactors}
\label{sec:Reactor}

The KamLAND 1~kton liquid scintillator detector located at Kamioka measured the neutrino flux from 53 power reactors corresponding to
701~Joule/cm${}^{2}$ \cite{Araki:2004mb}. An event rate of $365.2\pm23.7$ above 2.6~MeV for an 
exposure of 766~ton year from the
nuclear reactors was expected. The observed rate was 258 events
with a total background of $17.8\pm7.3$. The significant deficit,
interpreted in terms of neutrino oscillations, enables a measurement
of $\theta_{12}$, the neutrino 1-2 family mixing angle
($\sin^2\theta_{12} = 0.31^{+0.02}_{-0.03}$) as well as the mass
squared difference $\Delta m^2_{12} = (7.9\pm0.3)~\times 10^{-5}$eV${}^2$.

Future precision measurements are currently being investigated. Running KamLAND
for 2-3 more years would gain 30\% (4\%) reduction in the spread of
$\Delta m^2_{12}$ ($\theta_{12}$). Although it has been shown in Sections \ref{sec:SN} and \ref{sec:Geo}
that $\bar{\nu}_e$ originated from nuclear reactors can be a serious
background for diffuse supernova neutrino and geo-neutrino detection,
the Fréjus site can take benefit of the nuclear reactors located in
the Rh\^one valley to measure $\Delta m_{21}^2$ and $\sin^2\theta_{12}$.
In fact, approximately 67\% of the total reactor
$\bar{\nu}_e$ flux at Fréjus originates from four nuclear power plants
in the Rhone valley, located at distances between 115~km and 160~km. 
The indicated baselines are particularly suitable for
the study of the $\bar{\nu}_e$ oscillations driven by $\Delta m_{21}^2$. 
The authors of \cite{Petcov:2006gy} have investigated the possibility of using
one module of MEMPHYS (147~kton fiducial mass) 
doped with Gadolinium or the LENA detector, updating the previous work of \cite{Choubey:2004bf}. 
Above 3~MeV (2.6~MeV) the event rate is $59\ 980$ ($16\ 670$) events/year for
MEMPHYS (LENA), which is 2 orders of magnitude larger than the
KamLAND event rate.  
  
\begin{figure}
\begin{center}
\includegraphics[width=0.7\columnwidth]{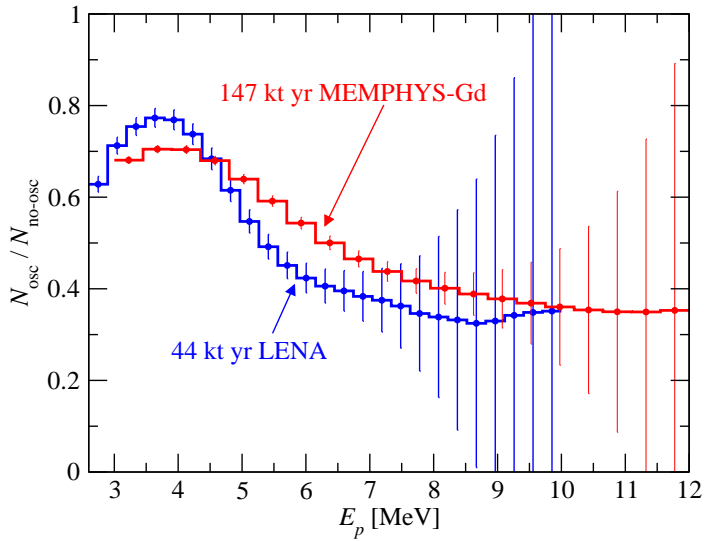}
\end{center}
  \caption{The ratio of the event spectra in positron energy 
  in the case of oscillations with $\Delta m_{21}^2 = 7.9\times 10^{-5}$~eV$^2$ and
  $\sin^2\theta_{12} = 0.30$ and in the absence of oscillations, 
  determined using one year data of MEMPHYS-Gd and LENA located at Frejus. 
  The error bars correspond to $1\sigma$ statistical error. Reprinted figure with permission from~\cite{Petcov:2006gy}.}
\label{fig:LENAMEMPHYS-reac-histo}
\end{figure}

\begin{figure}
\begin{center}
\includegraphics[width=0.7\columnwidth]{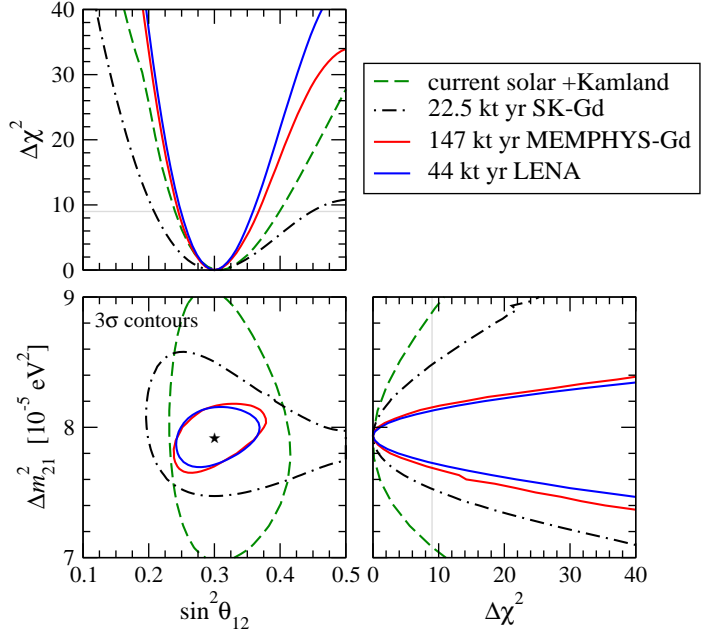}
\end{center}
  \caption{Accuracy of the determination of $\Delta m_{21}^2$ and
  $\sin^2\theta_{12}$, for one year data taking
  of MEMPHYS-Gd and LENA at Frejus, and Super-Kamiokande-Gd,
  compared to the current precision from solar neutrino and KamLAND
  data. The allowed regions at $3\sigma$ (2 d.o.f.) in the
  $\Delta m_{21}^2-\sin^2\theta_{12}$ plane, as well as the projections of the
  $\chi^2$ for each parameter are shown. Reprinted figure with permission from ~\cite{Petcov:2006gy}.}
\label{fig:reactor-sensitivities}
\end{figure}

In order to test the sensitivity of the experiments, the prompt energy 
spectrum is subdivided into 20 bins between 3~MeV
and 12~MeV for MEMPHYS-Gd and Super-Kamiokande-Gd, and into 25 bins between 2.6~MeV and
10~MeV for LENA (\refFig{fig:LENAMEMPHYS-reac-histo}). 
A $\chi^2$ analysis taking into account the statistical and systematical errors shows that each of the two
detectors, MEMPHYS-Gd and LENA if placed at Fréjus, can be exploited to yield a
precise determination of the solar neutrino oscillation
parameters $\Delta m_{21}^2$ and $\sin^2\theta_{12}$.  Within one year, the
3$\sigma$ uncertainties on $\Delta m_{21}^2$ and $\sin^2\theta_{12}$ can be
reduced respectively to less than 3\% and to approximately 20\% (\refFig{fig:reactor-sensitivities}). 
In comparison, the Gadolinium doped Super-Kamiokande detector that might be envisaged in a near future would reach
a similar precision only with a much longer data taking time.
Several years of reactor $\bar{\nu}_e$ data collected by 
MEMPHYS-Gd or LENA would allow a determination 
of $\Delta m_{21}^2$ and $\sin^2\theta_{12}$ with
uncertainties of approximately 1\% and 10\% at 3$\sigma$, respectively. 

\begin{figure}
\begin{center}
\includegraphics[width=0.7\columnwidth]{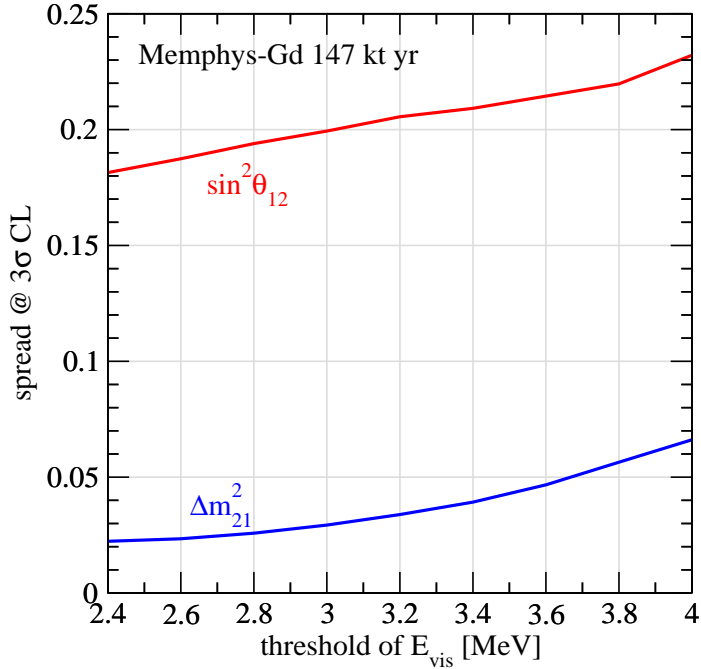}
\end{center}
  \caption{The accuracy of the determination of $\Delta m_{21}^2$ and
  $\sin^2\theta_{12}$, which can be obtained using one year of data
  from MEMPHYS-Gd as a function of the prompt energy threshold.}
\label{fig:reactor-MEMPHYS-threshold}
\end{figure}

However, some caveat are worth to be mentioned. The prompt energy trigger of 3~MeV requires a very low PMT dark 
current rate in the case of the MEMPHYS detector. If the energy threshold is higher,  the parameter precision decreases as can 
be seen in \refFig{fig:reactor-MEMPHYS-threshold}. The systematic uncertainties are also an 
important factor in the experiments under consideration, especially the determination of the
mixing angle, as those on the energy scale and the overall normalization.

Anyhow, the accuracy in the knowledge of the solar neutrino oscillation parameters, which can be
obtained in the high statistics experiments considered here, are
comparable to those that can be reached for the atmospheric neutrino
oscillation parameters $\Delta m_{31}^2$ and $\sin^2\theta_{23}$ with the future
long-baseline Super beam experiments such as T2HK or T2KK \cite{Ishitsuka:2005qi} in Japan, or SPL from
CERN to MEMPHYS. Hence, such reactor measurements would complete the
program of the high precision determination of the leading neutrino
oscillation parameters. 

%

\section{Neutrinos from particle accelerator beams}
\label{sec:oscillation}
Although the main physics goals of the proposed liquid-based detectors will be in the domain
of astro-particle physics, it would be economical and also very interesting from the physics point of view,
considering their possible use as "far" detectors for the future
neutrino facilities planned or under discussion in Europe, also given the large financial investment represented by
the detectors.
In this Section we review the physics program of the proposed observatories when using different accelerator 
neutrino beams. The main goals will be pushing the search for a non-zero (although very small) $\theta_{13}$ angle 
or its measurement in the case of a discovery previously made by one of the planned reactor or accelerator experiments
(Double-CHOOZ or T2K); searching for possible leptonic CP violation ($\delCP$); 
determining the mass hierarchy (the sign of $\Delta m^2_{31}$) and the $\theta_{23}$ octant 
($\theta_{23}>45^\circ$ or $\theta_{23}<45^\circ$). 
For this purpose we consider here 
the potentiality of a liquid Argon detector in an upgraded version of the existing CERN to Gran Sasso (CNGS) neutrino
beam, and of the MEMPHYS detector at the Fréjus using a possible new CERN proton driver (SPL) to upgrade to 4 MW the 
conventional neutrino beams (Super Beams). Another scheme contemplates a pure electron- (anti)neutrino production 
by radioactive ion decays (Beta Beam). Note that LENA is also a good candidate detector for the latter beam option. 
Finally, as an ultimate beam facility, one may think of producing very intense neutrino beams by means of 
muon decays (Neutrino Factory) that may well be detected with a liquid Argon detector such as GLACIER.  

The determination of the missing $U_{e3}$ ($\theta_{13}$ ) element of the neutrino mixing matrix is possible via the detection of
$\nu_\mu\rightarrow\nu_e$ oscillations at a baseline $L$ and energy $E$ given by the atmospheric neutrino signal, 
corresponding to a mass squared difference $E/L \sim \Delta m^2\simeq 2.5\times 10^{-3}\ eV^2$. 
The current layout of the CNGS beam from CERN to the Gran Sasso Laboratory has been optimized for a 
$\tau$-neutrino appearance search to be performed by the OPERA experiment \cite{Acquafredda:2006ki}.
This beam configuration provides limited sensitivity to the measurement of $U_{e3}$. 

Therefore,  we discuss the physics potential 
of an intensity-upgraded and energy-reoptimized CNGS neutrino beam coupled to an off-axis GLACIER
detector \cite{Meregaglia:2006du}. This idea is based on the possible upgrade of the
CERN PS or on a new machine (PS+) to deliver protons of 50~GeV/c
with a power of 200~kW. Post acceleration to SPS energies followed 
by extraction to the CNGS target region should allow to reach MW power, with neutrino energies peaked around 2 GeV.
In order to evaluate the physics potential one assumes five years of
running in the neutrino horn polarity plus five additional years in
the anti-neutrino mode. A systematic error on the
knowledge of the $\nu_e$ component of 5$\%$ is assumed. Given the excellent $\pi^0$
particle identification capabilities of GLACIER, the contamination of $\pi^0$ is negligible.

\begin{figure}[p]
\begin{center}
\includegraphics[width=0.7\columnwidth]{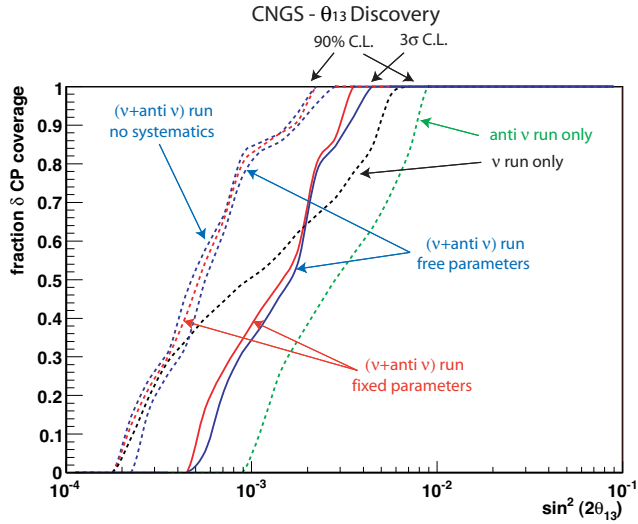}
\end{center}
\caption{\label{fig:fract_disc_theta}
GLACIER in the upgraded CNGS beam. Sensitivity to the discovery of $\theta_{13}$:
fraction of $\delta_{CP}$ coverage as a function of $\sin^22\theta_{13}$. Reprinted figure with permission from~\cite{Meregaglia:2006du}.}
\end{figure}
\begin{figure}[p]
\begin{center}
\includegraphics[width=0.7\columnwidth]{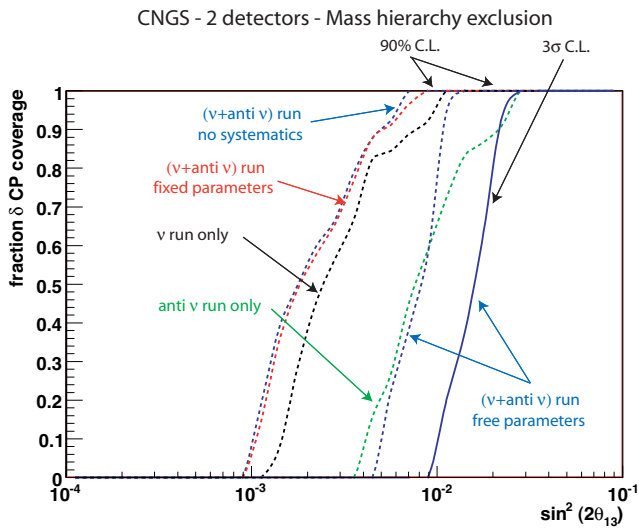}
\end{center}
\caption{\label{fig:fract_disc_dm}
Upgraded CNGS beam: mass hierarchy determination for a two detector configuration at
baselines of 850~km and 1050~km. Reprinted figure with permission from~\cite{Meregaglia:2006du}.}
\end{figure}

An off-axis beam search for $\nu_e$ appearance is performed with the
GLACIER detector located at 850 km from CERN. For an off-axis angle of
0.75$^o$,  $\theta_{13}$ can be discovered for full $\delta_{CP}$ coverage for $\sin^22\theta_{13}>0.004$ at
$3\sigma$ (Fig.~\ref{fig:fract_disc_theta}). 
At this rather modest baseline, the effect of CP violation and matter effects 
cannot be disentangled. In fact, the determination of the mass hierarchy
with half-coverage (50$\%$) is reached only for $\sin^22\theta_{13}>0.03$ at
$3\sigma$. A longer baseline (1050~km) and a larger off-axis angle
(1.5$^o$) would allow the detector to be sensitive to the first minimum and the second
maximum of the oscillation. This is the key to resolve the issue of mass
hierarchy. With this detector configuration, full coverage 
for $\delta_{CP}$ to determine the mass
hierarchy can be reached for $\sin^22\theta_{13}>0.04$ at
$3\sigma$. The sensitivity to mass hierarchy determination can be
improved by considering two off-axis detectors: one of 30 kton at 850
km and off-axis angle 0.75$^o$, a second one of 70 kton at 1050 km and
1.5$^0$ off-axis. Full coverage  for $\delta_{CP}$ to determine the mass
hierarchy can be reached for $\sin^22\theta_{13}>0.02$ at
$3\sigma$ (Fig.~\ref{fig:fract_disc_dm}). 
This two-detector configuration reaches very similar sensitivities to the ones of the T2KK proposal \cite{Ishitsuka:2005qi}.

\begin{figure}
\begin{center}
  \includegraphics[width=0.7\columnwidth]{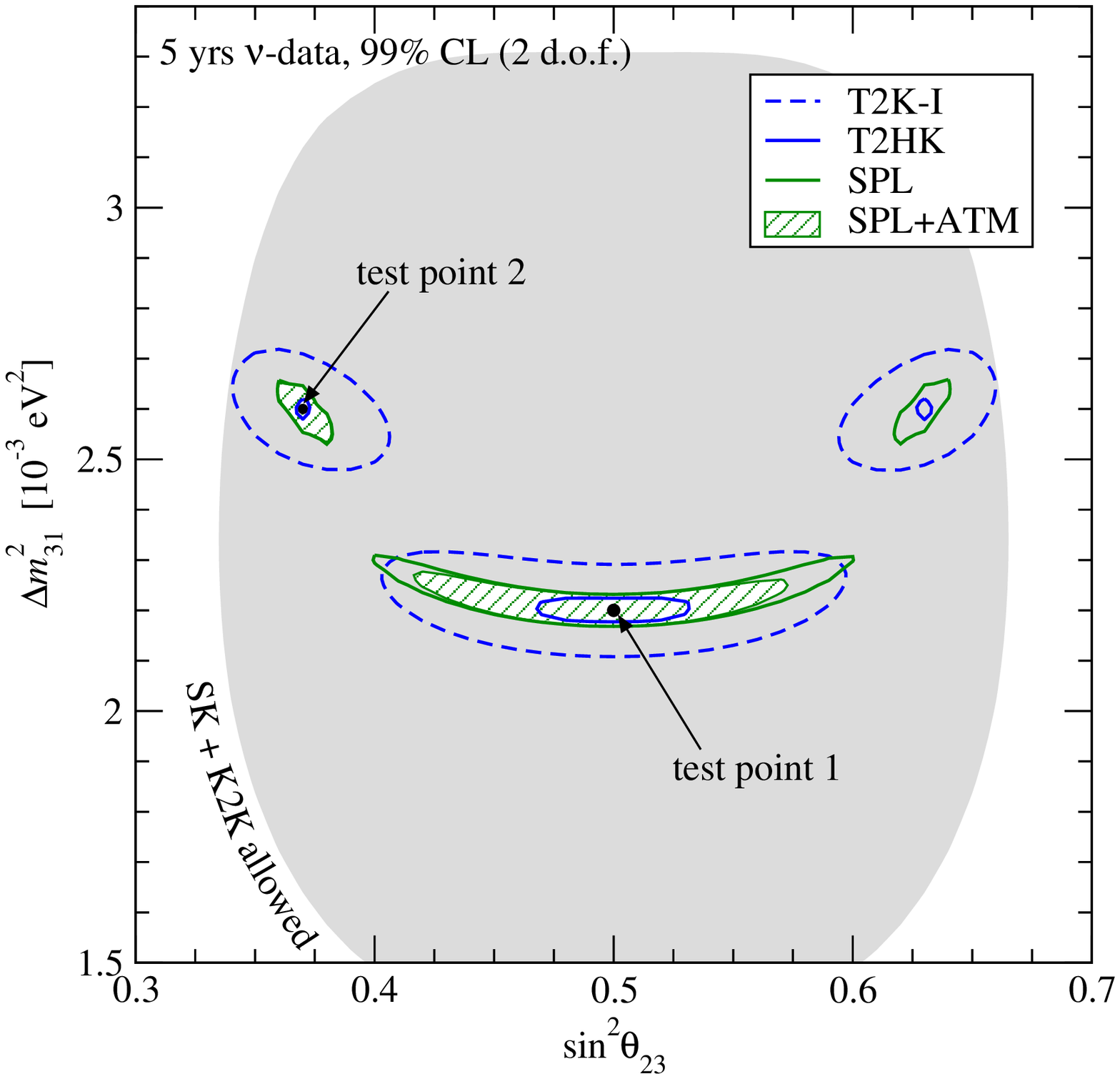}
\end{center}
  \caption{\label{fig:Phys-SPL-atm-params} Allowed regions of $\Delta
  m^2_{31}$ and $\sin^2\theta_{23}$ at 99\%~C.L. (2 d.o.f.)  after 5~years
  of neutrino data taking for ATM+SPL, T2K phase~I, ATM+T2HK, and the
  combination of SPL with 5~years of atmospheric neutrino data in the
  MEMPHYS detector. For the true parameter values we use $\Delta
  m^2_{31} = 2.2\, (2.6) \times 10^{-3}~\mathrm{eV}^2$ and
  $\sin^2\theta_{23} = 0.5 \, (0.37)$ for the test point 1 (2), and
  $\theta_{13} = 0$ and the solar parameters as: $\Delta m^2_{21} = 7.9 \times 10^{-5}~\mathrm{eV}^2$,
  $\sin^2\theta_{12} = 0.3$. The shaded region corresponds to the
  99\%~C.L. region from present SK and K2K data~\cite{Maltoni:2004ei}. Reprinted figure with permission from~\cite{Campagne:2006yx}.}
\end{figure}

Another notable possibility is the CERN-SPL Super Beam project.  
It is a conventional neutrino beam featuring a 4 MW SPL (Super-conducting Proton Linac) \cite{Gerigk:2006qi}
driver delivering protons onto a liquid Mercury target to generate 
an intense $\pi^+$ ($\pi^-$) beam with small contamination of kaons. 
The use of near and far detectors will allow both $\nu_{\mu}$ disappearance and
 $\nu_{\mu} \rightarrow \nu_e$ appearance studies.
The physics potential of the SPL Super Beam with MEMPHYS has been extensively studied \cite{Campagne:2006yx,Baldini:2006fi,ISS06}. However, the beam simulations will need some retuning after the forthcoming results of the CERN HARP 
experiment \cite{Catanesi:2001gi} on hadro-production.
 
After 5 years exposure in $\nu_\mu$ disappearance mode, a $3\sigma$ accuracy of (3-4)\% 
can be achieved on $\Delta m^2_{31}$, and an accuracy of 22\% (5\%) on $\sin^2\theta_{23}$ if the true value is $0.5$ (0.37), namely in case of maximal or non-maximal mixing (\refFig{fig:Phys-SPL-atm-params}). The use of atmospheric neutrinos can contribute to solving
the octant ambiguity in case of non-maximal mixing as it is shown in \refFig{fig:Phys-SPL-atm-params}. Note however, 
that thanks to a higher energy beam ($\sim 750$~MeV), the T2HK project\footnote{Here, we  to the project where a 
4 MW proton driver is built at KEK to deliver an intense neutrino beam detected by a large \WC\ detector.} can benefit from a much lower dependence on the Fermi motion to obtain a better energy resolution.

\begin{figure}[p]
\begin{center}
  \includegraphics[width=0.7\columnwidth]{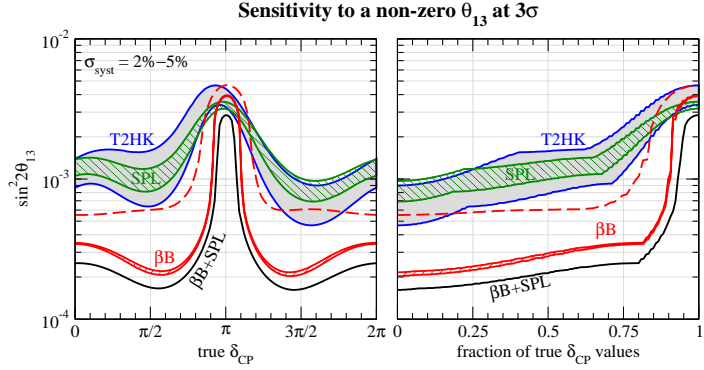}
\end{center}
  \caption{$3\sigma$ discovery sensitivity to $\stheta$ for
  Beta Beam, SPL, and T2HK as a function of the true value of \delCP\
  (left panel) and as a function of the fraction of all possible
  values of \delCP\ (right panel). The width of the bands corresponds
  to values for the systematical errors between 2\% and 5\%. The
  dashed curve corresponds to the Beta Beam sensitivity with the fluxes reduced by a factor 2. Reprinted figure with permission from~\cite{Campagne:2006yx}.\label{fig:Phys-SPLBB-th13}}
\end{figure}

\begin{figure}[p]
\begin{center}
   \includegraphics[width=0.7\columnwidth]{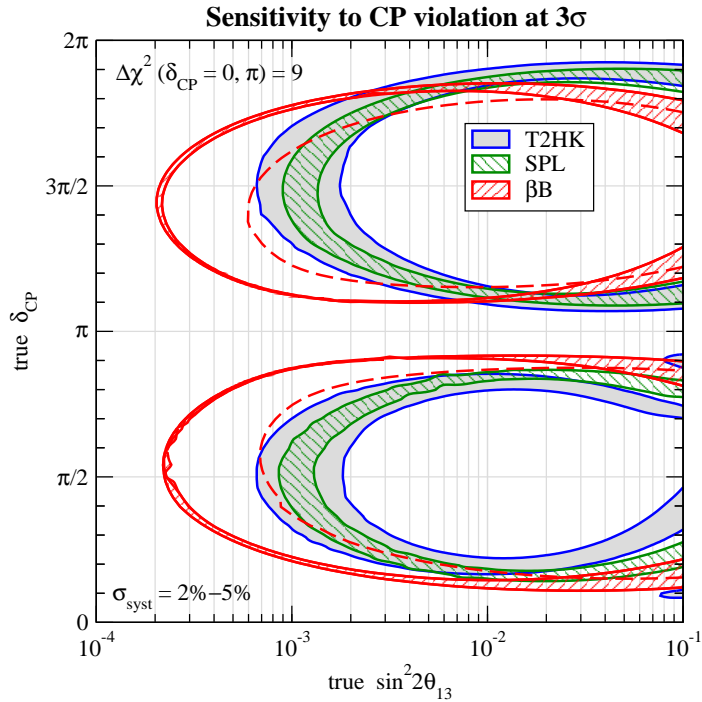}
\end{center}
   \caption{CP violation discovery potential for Beta Beam, SPL, and T2HK: For
   parameter values inside the ellipse-shaped curves CP conserving
   values of \delCP\ can be excluded at $3\sigma$ $(\Delta\chi^2>9)$.
   The width of the bands corresponds to values for the systematic
   errors from 2\% to 5\%. The dashed curve is described in \refFig{fig:Phys-SPLBB-th13}. Reprinted figure with permission from~\cite{Campagne:2006yx}.
   \label{fig:Phys-SPLBB-CPV}}
\end{figure}

In appearance mode (2 years $\nu_\mu$ plus
8 years \nubarmu), a $3\sigma$ discovery of non-zero $\theta_{13}$, irrespective of the actual true value of $\delCP$, is achieved 
for $\stheta\gtrsim 4\ 10^{-3}$ ($\thetaot \gtrsim 3.6^\circ$) as shown in \refFig{fig:Phys-SPLBB-th13}. For maximal CP violation 
($\delCP^\mathrm{true} = \pi/2, \, 3\pi/2$) the same discovery level can be achieved for $\stheta\gtrsim 8\ 10^{-4}$ 
($\thetaot \gtrsim 0.8^\circ$). The best sensitivity for testing CP violation ($i.e$ the data cannot be fitted with $\delCP =0$ nor $\delCP=\pi$) is achieved for $\stheta\approx 10^{-3}$ ($\thetaot \approx 0.9^\circ$) as shown in \refFig{fig:Phys-SPLBB-CPV}. The maximum sensitivity is achieved for $\stheta\sim 10^{-2}$ where the CP violation can be established at 3$\sigma$ for 73\% of all the $\delCP^\mathrm{true}$.

Although quite powerful, the proposed SPL Super Beam is a conventional neutrino beam with known limitations due to the low 
production rate of anti-neutrinos compared to neutrinos which, in addition to a smaller charged-current cross-section, 
imposes to run 4 times longer in anti-neutrino mode, and implies difficulty to set up an accurate beam simulation, and to 
design a non-trivial near detector setup mastering the background level. Thus, a new type of neutrino beam, the so-called Beta Beam is being considered. 
The idea is to generate pure, well collimated and intense \nue  (\nubare) beams by producing, collecting, and accelerating 
radioactive ions \cite{Zucchelli:2002sa}.
The resulting Beta Beam  spectra can be easily computed knowing the beta-decay spectrum of the parent
ion and the Lorentz boost factor $\gamma$, and these beams are virtually free from other 
background flavors. The best ion candidates so far are  $^{18}$Ne  and $^6$He for \nue and \nubare,  respectively.
A baseline study for the Beta Beam has been initiated at CERN, and is now going on within the European FP6 design study for EURISOL. 

The potential of such Beta Beam sent to MEMPHYS has been studied in the context of the baseline scenario, using reference fluxes of $5.8 \times 10^{18}$ \He\ useful decays/year and $2.2 \times10^{18}$ \Ne\  decays/year, corresponding to a
reasonable estimate by experts in the field of the ultimately
achievable fluxes.  The optimal values is actually $\gamma = 100$
for both species, and the corresponding performance have been recently reviewed in \cite{Campagne:2006yx,Baldini:2006fi,ISS06}.

In Figs.~\ref{fig:Phys-SPLBB-th13},\ref{fig:Phys-SPLBB-CPV} the results of running a Beta Beam during 10 years (5 years with neutrinos and 5 years with anti-neutrinos) is shown and prove to be far better compared to an SPL Super beam run, especially for maximal CP violation  where a non-zero $\thetaot$ value can be stated at $3\sigma$ for $\stheta\gtrsim 2\ 10^{-4}$ ($\thetaot \gtrsim 0.4^\circ$). Moreover, it is noticeable that the Beta Beam is less affected by systematic errors of the background compared to the SPL Super beam and T2HK.

Before combining the two possible CERN beam options, relevant for the proposed European underground observatories,
let us consider LENA as potential detector. LENA, with a fiducial volume of $\sim 45$~kton, can as well be used as
detector for a low-energy Beta Beam oscillation experiment. In the energy
range $0.2-1.2$~GeV, the performed simulations show that muon events are 
separable from electron events due to their different track
lengths in the detector and due to the electron emitted in the muon decay.
For high energies, muons travel longer than electrons, as the latter undergo scattering and bremsstrahlung. This results in different
distributions of the number of photons and the timing pattern, which can be used to distinguish between the two classes of events. For low energies, muons can be recognized by observing the electron of its
succeeding decay after a mean time of 2.2~$\mu$s. By using both criteria, an efficiency of $\sim 90$~\% for muon appearance
has been calculated with acceptance of 1~\% electron background. The advantage of using a liquid scintillator detector for such an
experiment is the good energy reconstruction of the neutrino beam.
However, neutrinos of these energies can produce $\Delta$ resonances
which subsequently decay into a nucleon and a pion. In \WC\ detectors,
pions with energies under the Cherenkov threshold contribute to the
uncertainty of the neutrino energy. In LENA these particles can be
detected. The effect of pion production and similar reactions is currently under investigation in order to estimate the actual energy
resolution.

We also mention a very recent development of the Beta Beam concept \cite{Rubbia:2006pi} 
based on a very promising alternative for the 
production of ions and on the possibility of having monochromatic, single-flavor neutrino beams
by using ions decaying through the electron capture process \cite{Bernabeu:2005jh,Sato:2005ma}.
In particular, such beams would be suitable to precisely measure neutrino cross-sections in a near detector with the
possibility of an energy scan by varying the $\gamma$ value of the ions.
Since a Beta Beam uses only a small fraction of the protons available from the
SPL, Super and Beta Beams can be run at the same time. The combination of a Super Beam and a Beta Beam 
offers advantages from the experimental point of view since the
same parameters $\theta_{13}$ and $\delta_{CP}$ can be measured in many
different ways, using 2 pairs of CP related channels, 2 pairs of T related
channels, and 2 pairs of CPT related channels which should all give
coherent results. In this way, the estimates of systematic errors,
different for each beam, will be experimentally cross-checked.
Needless to say, the unoscillated data for a given beam will provide a large
sample of events corresponding to the small searched-for signal with the
other beam, adding more handles to the understanding of the detector
response.

\begin{figure}
\begin{center}
   \includegraphics[width=0.7\columnwidth]{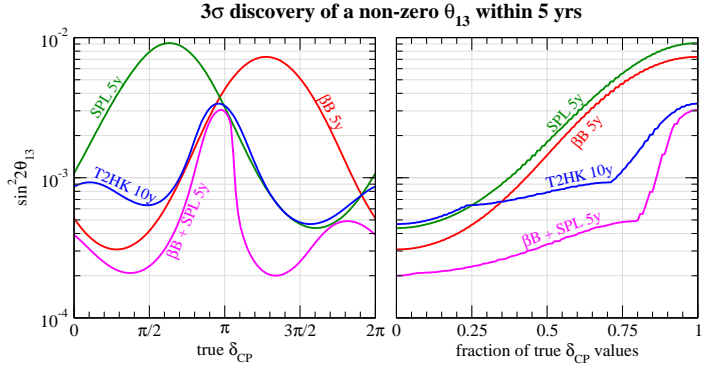}
\end{center}
   \caption{Discovery potential of a finite value of $\stheta$ at
   $3\sigma$ $(\Delta\chi^2>9)$ for 5~years neutrino data from
   Beta Beam, SPL, and the combination of Beta Beam + SPL compared to
   10~years data from T2HK (2~years neutrinos + 8~years antineutrinos). Reprinted figure with permission from~\cite{Campagne:2006yx}.
   \label{fig:Phys-SPLBB-th13-5years}}
\end{figure}

The combination of the Beta Beam and the Super Beam
will allow to use neutrino modes only: $\nu_\mu$ for SPL and $\nu_e$ for Beta Beam. 
If CPT symmetry is assumed, all the information can be 
obtained as $P_{\bar\nu_e\to\bar\nu_\mu} = P_{\nu_\mu\to\nu_e}$ and $P_{\bar\nu_\mu\to\bar\nu_e} = P_{\nu_e\to\nu_\mu}$. We illustrate this synergy in \refFig{fig:Phys-SPLBB-th13-5years}. In this scenario, time consuming anti-neutrino running can be avoided keeping the same physics discovery potential. 

One can also combine SPL, Beta Beam and the atmospheric neutrino experiments to reduce the
parameter degeneracies which lead to disconnected regions on the multi-dimensional space of oscillation parameters.
One can look at \cite{Burguet-Castell:2001ez,Minakata:2001qm,Fogli:1996pv} for the definitions of {\it intrinsic}, {\it hierarchy}, and {\it octant} degeneracies. As we have seen above, atmospheric neutrinos, mainly multi-GeV $e$-like events, are sensitive to the
neutrino mass hierarchy if $\theta_{13}$ is sufficiently large due to
Earth matter effects, whilst sub-GeV $e$-like events provide sensitivity to the octant of
$\theta_{23}$ due to oscillations with $\Delta m^2_{21}$.

\begin{figure}
\begin{center}
\includegraphics[width=0.7\columnwidth]{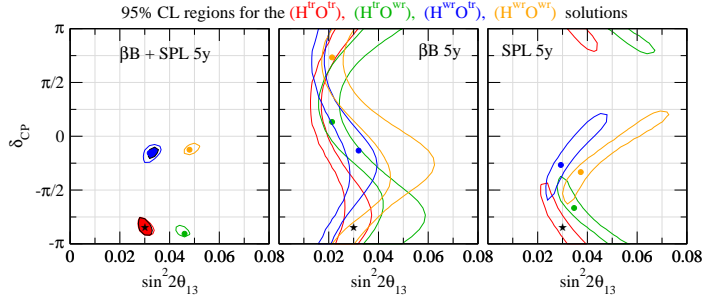}
\end{center}
  \caption{Allowed regions in $\sin^22\theta_{13}$ and
  $\delta_{CP}$ for 5~years data (neutrinos only) from Beta Beam,
  SPL, and the combination. $\mathrm{H^{tr/wr} (O^{tr/wr})}$ refers to
  solutions with the true/wrong mass hierarchy (octant of
  $\theta_{23}$). For the colored regions in the left panel also
  5~years of atmospheric data are included; the solution with the
  wrong hierarchy has $\Delta\chi^2 = 3.3$. The true parameter
  values are $\delta_{CP} = -0.85 \pi$, $\sin^22\theta_{13} =
  0.03$, $\sin^2\theta_{23} = 0.6$. For the Beta Beam
  only analysis (middle panel) an external accuracy of 2\% (3\%) for
  $|\Delta m^2_{31}|$ ($\theta_{23}$) has been assumed, whereas for
  the left and right panel the default value of 10\% has been used. Reprinted figure with permission from~\cite{Campagne:2006yx}.}
\label{fig:Phys-SPLBB-degeneracies_5years}
\end{figure}

The result of running during 5 years in neutrino mode for SPL and Beta Beam, adding further the 
atmospheric neutrino data, is shown in \refFig{fig:Phys-SPLBB-degeneracies_5years} \cite{Campagne:2006yx}. 
One can appreciate that practically all degeneracies can be eliminated as only the solution with the wrong sign 
survives with a $\Delta \chi^2 = 3.3$. 
This last degeneracy can be completely eliminated by using a neutrino running mode combined with anti-neutrino mode and ATM 
data \cite{Campagne:2006yx}. However, the example shown is a favorable case with $\sin^2\theta_{23}=0.6$ and in general, 
for $\sin^2\theta_{23}<0.5$, the impact of the atmospheric data is weaker. 
So, as a generic case, for the CERN-MEMPHYS project, one is left with the four intrinsic degeneracies. 
However, the important observation in \refFig{fig:Phys-SPLBB-degeneracies_5years} is that
degeneracies have only a very small impact on the CP violation discovery, in the sense that if the true solution is CP violating also
the fake solutions are located at CP violating values of
$\delCP$. Therefore, thanks to the relatively short baseline without matter effect, even if degeneracies
affect the precise determination of $\theta_{13}$ and $\delCP$, they
have only a small impact on the CP violation discovery potential. Furthermore, one would quote explicitly the four possible sets of parameters with their respective confidential level. It is also clear from the figure that the sign($\Delta
m^2_{31}$) degeneracy has practically no effect on the $\theta_{13}$
measurement, whereas the octant degeneracy has very little impact on the determination of $\delCP$.

Some other features of the atmospheric neutrino data are presented in \refSec{sec:Phys-Atm-neut}.
In order to fully exploit the possibilities offered by a Neutrino
Factory, the detector should be capable of identifying  and measuring all three charged lepton flavors
produced in charged-current interactions and of measuring 
their charges in order to identify the incoming neutrino helicity. 
The GLACIER concept in its non-magnetized option provides a background-free identification of electron-neutrino charged-current events and a kinematical selection of tau-neutrino charged-current interactions.
We can assume that charge discrimination is available for muons reaching an external magnetized-Fe spectrometer.

Another interesting and extremely challenging possibility would consist in magnetizing the whole 
liquid Argon volume \cite{Badertscher:2005te,Ereditato:2005yx}. This set-up would allow the clean classification of events 
into electrons, right-sign muons, wrong-sign muons and no-lepton categories.
In addition, high granularity permits a clean detection of quasi-elastic events, which
provide a selection of the neutrino electron helicity by detecting the final state proton,
without the need of an electron charge measurement. 
Table~\ref{tab:rates} summarizes the expected rates for GLACIER and $10^{20}$ muon decays at a neutrino factory with stored muons 
having an energy of 30 GeV \cite{Bueno:2000fg}.  $N_{tot}$ is the total number of events and $N_{qe}$ is the number
of quasi-elastic events. 

\begin{table}
\caption{\label{tab:rates}Expected events rates for GLACIER in a Neutrino Factory beam,
assuming no oscillations and for $10^{20}$ muon decays (E$_\mu$=30 GeV).  
$N_{tot}$ is the total number of events and $N_{qe}$ is the number of quasi-elastic events.}
\lineup
\begin{tabular}{@{}llllllll}
\br
\multicolumn{8}{@{}c}{Event rates for various baselines} \\ 
\mr
 & & \multicolumn{2}{@{}c}{$L=732$~km} & \multicolumn{2}{c}{$L=2900$~km} & 
\multicolumn{2}{@{}c}{$L=7400$~km} \\
 & & $N_{tot}$ & $N_{qe}$ & $N_{tot}$ & $N_{qe}$ & $N_{tot}$ & $N_{qe}$ \\
 & $\numu$ CC & 2260\ 000 & 90\ 400 & 144\ 000 & 5760 & 22\ 700 & 900 \\
$\mu^-$ & $\numu$ NC & \phantom{0}673\ 000 & --- &  \phantom{0}41\ 200 & --- & \phantom{0}\ 6800 & ---  \\
$10^{20}$ decays & $\anue$ CC &  \phantom{0}871\ 000 & 34\ 800 & \phantom{0}55\ 300 & 2200 & \phantom{0}\ 8750 & 350 \\
 & $\anue$ NC & \phantom{0}302\ 000 & ---  & \phantom{0}19\ 900 & ---  &  \phantom{0}\ 3000 & ---  \\ \mr
 & $\anumu$ CC & 1010\ 000 & 40\ 400 & \phantom{0}63\ 800 & 2550 & 10\ 000 & 400 \\
$\mu^+$ & $\anumu$ NC &  \phantom{0}353\ 000 & --- & \phantom{0}22\ 400 & --- &  \phantom{0}\ 3500 & --- \\
$10^{20}$ decays & $\nue$ CC &  1970\ 000 & 78\ 800 & 129\ 000 & 5160 & 19\ 800 & 800 \\
 & $\nue$ NC &  \phantom{0}579\ 000 & --- & \phantom{0}36\ 700 & --- &  \phantom{0}\ 5800 & --- \\
 \br
\end{tabular}
\end{table}

\begin{figure}
\begin{center}
\includegraphics[width=0.7\columnwidth]{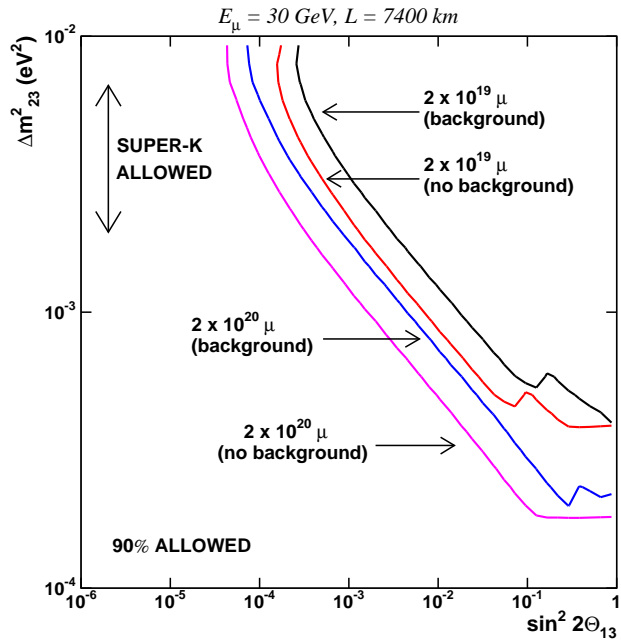}
\end{center}
\caption{\label{fig:t13sensitivity} GLACIER sensitivity to the measurement of $\theta_{13}$. Reprinted figure with permission from~\cite{Bueno:2000fg}.}
\end{figure}

Figure~\ref{fig:t13sensitivity} 
shows the expected sensitivity in the measurement of $\theta_{13}$ 
for a baseline of  7400 km. The maximal sensitivity to $\theta_{13}$ is achieved for very small
background levels, since one is looking in this case for small signals; most of the information is coming from the clean
wrong-sign muon class and from quasi-elastic events.  On the other hand,  if its value is not too small, for a 
measurement of $\theta_{13}$, the signal/background ratio could be not so crucial, and also the other event classes can contribute to this measurement.

A Neutrino Factory should aim to over-constrain the oscillation pattern, in order to look for
unexpected new physics effects. This can be achieved in global fits of the parameters, where the unitarity of the mixing matrix is 
not strictly assumed. Using a detector able to identify the $\tau$ lepton production via
kinematic means, it is possible to verify the unitarity in 
$\nu_\mu\to\nu_\tau$ and $\nu_e\to\nu_\tau$ transitions. 

The study of CP violation in the lepton system probably is the most ambitious goal of  an experiment at a Neutrino Factory. 
Matter effects can mimic CP violation; however, a multi-parameter fit
at the right baseline can allow a simultaneous determination of
matter and CP violating parameters. To detect CP violation effects, the most favorable choice of 
neutrino energy $E_\nu$ and baseline $L$ is in the region of  the first maximum, given by $(L/E_\nu)^{max}\simeq 500$ km/GeV 
for $|\Delta m^2_{32}|=2.5\times 10^{-3}\rm\ eV^2$ \cite{Bueno:2001jd}. 
To study oscillations in this region, one has to require that the energy of the "first-maximum'' be smaller than 
the MSW resonance energy: $2\sqrt{2}G_Fn_eE^{max}_\nu\lesssim\Delta m^2_{32}\cos 2\theta_{13}$. 
This fixes a limit on the baseline $L_{max} \approx 5000$~km 
beyond which matter effects spoil the sensitivity.

\begin{figure}
\begin{center}
\includegraphics[width=0.7\columnwidth]{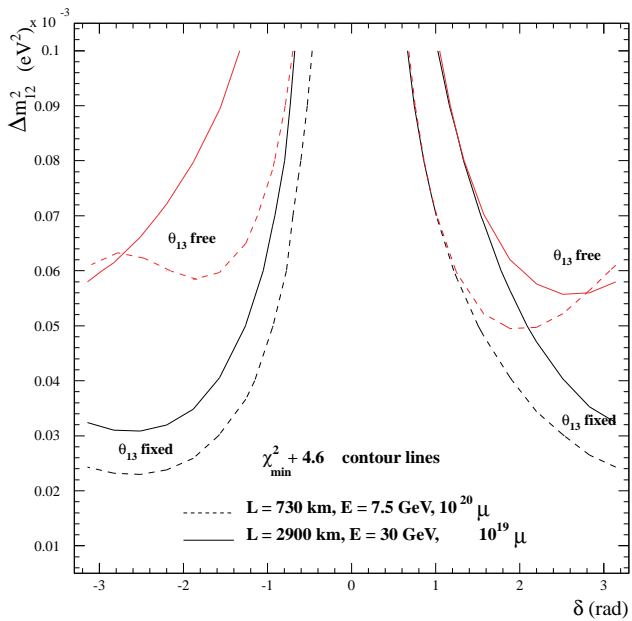}
\end{center}
\caption{\label{fig:cpsensitivity} GLACIER 90\% C.L. sensitivity on the $CP$-phase $\delta_{CP}$ as a function of
$\Delta m^2_{21}$ for the two considered baselines.
The reference oscillation parameters are
$\Delta m^2_{32}=3\times 10^{-3}\ \rm eV^2$,
$\sin^2 \theta_{23} = 0.5$,
$\sin^2 \theta_{12} = 0.5$,
$\sin^2 2\theta_{13} = 0.05$ and
$\delta_{CP} = 0$.
The lower curves are made fixing all parameters to the reference values
while for the upper curves $\theta_{13}$ is free. Reprinted figure with permission from~\cite{Bueno:2001jd}.}
\end{figure}

As an example, \refFig{fig:cpsensitivity} shows the sensitivity 
to the CP violating phase $\delta_{CP}$ for two concrete cases. 
The events are classified in the five categories previously mentioned, 
assuming an electron charge confusion of 0.1$\%$. The exclusion 
regions in the $\Delta m^2_{12} - \delta_{CP}$ plane are determined by fitting the 
visible energy distributions, provided that the electron detection efficiency is $\sim 20\%$. The excluded regions 
extend up to values of $|\delta_{CP}|$ close to $\pi$,  even when $\theta_{13}$ is left free.

\section{Conclusions and outlook}
\label{sec:Phys-Summary}

In this paper we discuss the importance of outstanding
physics phenomena such as the possible instability of matter (proton decay), the production of neutrinos
in supernovae, in the Sun and in the interior of the Earth, as well as the recently discovered 
process of neutrino oscillations, also detectable through artificial neutrinos produced by nuclear reactors and
particle accelerators.

All the above physics subjects, key issues for particle physics, astro-particle physics, astrophysics and cosmology,
call for a new generation of multipurpose, underground observatories based on improved detection techniques. 

The envisioned detectors must necessarily be very massive (and consequently large) and 
able to provide very low experimental background. 
The required signal to noise ratio can only be achieved in underground laboratories suitably shielded against cosmic-rays
and environmental radioactivity. Some candidate sites in Europe have been identified and we are progressing
in assessing in detail their capabilities.

We have identified three different and, to a large extent, complementary technologies capable of meeting the challenge, based
on large scale use of liquids for building large-size, volume-instrumented detectors.
The three proposed large-mass, liquid-based
detectors for future underground observatories for particle physics in Europe (GLACIER, LENA and MEMPHYS),
although based on completely different detection techniques 
(liquid Argon, liquid scintillator and \WC), share a similar, very rich physics program. For some cases of interest their 
detection properties are quite complementary.  
A summary of the scientific case presented in this paper is given for astro-particle physics topics 
in Table \ref{tab:Phys-potential-summary1}.

\begin{sidewaystable}
\caption{\label{tab:Phys-potential-summary1}
Summary of the physics potential of the proposed detectors for astro-particle physics topics.  The (*) stands for the case where 
Gadolinium salt is added to the water of one of the MEMPHYS shafts.}
\begin{indented}
\item[]
\begin{tabular}{@{}llll}
\br
Topics             &       GLACIER            &    {LENA}    &      {MEMPHYS}\\
                   &         100~kton                    &      50~kton        & 440~kton \\
\mr
\multicolumn{4}{@{}l}{{Proton decay}}  \\ 
$e^+\pi^0$ & 	$0.5\times 10^{35}$ & ---           &  $1.0\times 10^{35}$ \\
$\bar{\nu}K^+$  & 	$1.1\times 10^{35}$ & $0.4\times 10^{35}$            &  $0.2\times 10^{35}$ \\

\mr

\multicolumn{4}{@{}l}{{SN $\nu$ (10~kpc)}}          \\
CC & $2.5\times10^4 (\nue)$ & $9.0\times10^3 (\nubare)$ & $2.0\times10^5 (\nubare)$ \\
NC & $3.0\times10^4$ & $3.0\times10^3$ & --- \\
ES & $1.0\times10^3 (e)$ & $7.0\times10^3 (p)$ & $1.0\times10^3 (e)$ \\    
\mr

{DSNB $\nu$}

(S/B 5 years) & 40-60/30 & 9-110/7  & 43-109/47 (*) \\

\mr

\multicolumn{4}{@{}l}{{Solar $\nu$ (Evts. 1 year)}}  \\ 
$^8$B ES      & $ 4.5\times10^4$ & $1.6\times10^4$ & $1.1\times10^5$ \\
$^8$B CC     &           ---              & $360$           & ---\\
$^7$Be          &            ---             & $2.0\times10^6$ &  ---\\
$pep$             &              ---           & $7.7\times10^4$ &    ---\\
\mr

{Atmospheric $\nu$}
(Evts. 1 year)   &  $1.1\times10^4$                &     ---    &   $4.0\times10^4$ (1-ring only) \\  
\mr

{Geo $\nu$}
(Evts. 1 year)   &   below threshold                   &    $\approx 1000$ & need 2~MeV threshold \\
\mr

{Reactor $\nu$}
(Evts. 1 year))  &  ---                      &    $1.7\times10^4$        &  $6.0\times10^4$ (*) \\
\mr

{Dark Matter}
(Evts. 10 years)   &  \parbox[t]{4cm}{3 events\\ ($\sigma_{ES}=10^{-4}$,$M>20$~GeV)} & ---   & --- \\
\br
\end{tabular}
\end{indented}
\end{sidewaystable}
\ack

We wish to warmly acknowledge support from all the various funding agencies.  We wish to thank the EU framework 6 project ILIAS for providing assistance particularly regarding underground site aspects (contract 8R113-CT-2004-506222).

\newpage
\section*{References}
\bibliography{Campagne}
\end{document}